\documentclass[pra,10pt,twocolumn,superscriptaddress,showpacs]{revtex4-1}

\usepackage{lipsum} 
\usepackage{epsfig}
\usepackage{amsmath}
\usepackage{graphicx}
\usepackage{graphics}
\usepackage{float}
\usepackage{amssymb}
\usepackage[usenames,dvipsnames]{color}
\usepackage{natbib}
\usepackage{epstopdf}
\usepackage{verbatim}
\usepackage{wrapfig}
\usepackage{mathtools}
\usepackage{dsfont}

\usepackage{breqn}

\graphicspath{{Figure/}}

\usepackage{bm}

\makeatletter
\let\cat@comma@active\@empty
\makeatother

\newcommand{\bra}[1]{\left\langle #1\right|}
\newcommand{\ket}[1]{\left| #1\right\rangle}
\newcommand{\ip}[2]{\left\langle #1 | #2\right\rangle}

\newcommand{\opav}[3]{\langle #1 | #2 | #3 \rangle}

\newcommand{\beq}{\begin{equation}}
\newcommand{\eeq}{\end{equation}}
\newcommand{\tr}{\text{Tr}}

\begin{document}

\title{Geometric phase corrected by initial system-environment correlations}

\author{Sharoon Austin}
\affiliation{School of Science \& Engineering, Lahore University of Management Sciences (LUMS), Opposite Sector U, D.H.A, Lahore 54792, Pakistan}

\author{Sheraz Zahid}
\affiliation{School of Science \& Engineering, Lahore University of Management Sciences (LUMS), Opposite Sector U, D.H.A, Lahore 54792, Pakistan}

\author{Adam Zaman Chaudhry}
\email{adam.zaman@lums.edu.pk}
\affiliation{School of Science \& Engineering, Lahore University of Management Sciences (LUMS), Opposite Sector U, D.H.A, Lahore 54792, Pakistan}

\begin{abstract}

We find the geometric phase of a two-level system undergoing pure dephasing via interaction with an arbitrary environment, taking into account the effect of the initial system-environment correlations. We use our formalism to calculate the geometric phase for the two-level system in the presence of both harmonic oscillator and spin environments, and we consider the initial state of the two-level system to be prepared by a projective measurement or a unitary operation. The geometric phase is evaluated for a variety of parameters such as the  system-environment coupling strength to show that the initial correlations can affect the geometric phase very significantly even for weak and moderate system-environment coupling strengths. Moreover, the correction to the geometric phase due to the system-environment coupling generally becomes smaller (and can even be zero) if initial system-environment correlations are taken into account, thus implying that the system-environment correlations can increase the robustness of the geometric phase.

\end{abstract}


\maketitle

\section{Introduction}

The geometric phase is the phase information acquired by a system due to its cyclic evolution in a curved parameter space \cite{SjoqvistIntJQuantumChem2015,CohenNatureReview2019}. This phenomenon was first studied by Pancharatnam in optics \cite{Pancharatnam1956} and by Longuet-Higgins \cite{Higgins1975} and Stone \cite{Stone1976} in quantum chemistry. Berry's finding that the geometric phase arises generally in the study of closed quantum systems undergoing cyclic adiabatic evolutions ignited interest in the subject \cite{Berry1984}. Aharonov and Anand thereafter generalized the geometric phase to non-adiabatic evolutions, showing that the phase depends on the geometry of the path followed by the system in the projective Hilbert space \cite{AharonovPRL1987}, while Uhlmann considered the geometric phase for mixed quantum states \cite{UhlmannAnnPhys1989} which was further generalized by Sjoqvist \textit{et al.} \cite{SjoqvistPRL2000}. On the experimental front, the geometric phase has been observed in nuclear magnetic resonance \cite{SuterMolPhys1987}, superconducting \cite{LeekScience2007}, and optical setups \cite{SimonPRL1988}, amongst others.     

Besides its theoretical importance, the geometric phase has practical applications as well. In particular, due to its geometric nature, the geometric phase may have intrinsic resistance to external noise, which makes it an attractive tool for robust quantum information processing \cite{ZanardiPhysLettA1999,JonesNature2000,FalciNature2000,DuanScience2001,MatsumotoPRL2001,LiebfriedNature2003}. It is then important to extend the study of the geometric phase to open quantum systems where the effect of the environment on the geometric phase can be investigated. Different approaches have been used to investigate the effect of the environment on the geometric phase \cite{EricssonPRA2003, VedralPRL2003, TongPRL2004,WhitneyPRL2005,YiPRA2006,VillarPRA2006,DajkaJPhysA2008,VillarPRA2010,VillarPRL2010,VillarPRA2011,VillarPRA2013,VillarPRA2015}. In particular, emphasis has been on a single two-level system undergoing pure dephasing, that is, it is assumed that dephasing plays a much more dominant role compared to relaxation effects. In this case, starting from a product state of the two-level system and the environment in thermal equilibrium, the density matrix of the two-level system can be computed as a function of time, and the geometric phase can then be obtained. Of particular importance to us is Ref.~\cite{VillarPRA2015} where the effect of non-Markovianity on the geometric phase is studied. Given that memory effects can play a role, it is then natural to consider the effect of initial system-environment correlations on the geometric phase as well \cite{HakimPRA1985, HaakePRA1985, Grabert1988, SmithPRA1990, GrabertPRE1997, PazPRA1997, LutzPRA2003, BanerjeePRE2003, vanKampen2004, BanPRA2009, HanggiPRL2009, UchiyamaPRA2010, TanimuraPRL2010, SmirnePRA2010, DajkaPRA2010, ZhangPRA2010,TanPRA2011, CKLeePRE2012,MorozovPRA2012, SeminPRA2012,  ChaudhryPRA2013a,ChaudhryPRA2013b,ChaudhryCJC2013,FanchiniSciRep2014,FanSciRep2015correlation,ChenPRA2016,VegaRMP2017,VegaPRA2017,ShibataJPhysA2017,CaoPRA2017,MehwishEurPhysJD2019}. The effect of the initial correlations is expected to be especially significant if the system-environment coupling is not weak, since in this case, the initial state can no longer be assumed to be a product state of the system and the environment thermal equilibrium state. However, to date, to the best of our knowledge, the effect of the initial system-environment correlations on the geometric phase has not been studied. In this work, we aim to study the geometric phase for the pure dephasing model, taking the initial system-environment correlations into account. 

We start by deriving general expressions for the geometric phase of a two level system undergoing pure dephasing for both initially pure and mixed states. Our expressions are general in the sense that we do not make any assumptions regarding the form of the environment or the system-environment coupling, and they take the initial system-environment correlations into account. We then apply these expressions to two concrete well-known system-environment models: a two-level system undergoing dephasing via interaction with a harmonic oscillator environment, and a two-level system undergoing pure dephasing due to a spin environment. Both of these models are exactly solvable for arbitrary system-environment coupling strengths even if initial system-environment correlations are taken into account. The initial state of the two-level system is prepared either by performing a projective measurement on the system only (the initial state of the system is pure in this case), or by performing a unitary operation on the system (the initial state is now, in general, mixed). Using the exact solutions, we investigate the effect of the initial system-environment correlations on the geometric phase as various physical parameters such as the system-environment coupling strength and the temperature are varied. We find that, in general, the initial correlations can affect the geometric phase very significantly, even for weak and moderate system-environment coupling strengths. Interestingly, the initial correlations can make the geometric phase more robust; in fact, the correction to the geometric phase due to the environment can become zero for specific values of system-environment parameters if the initial correlations are taken into account. 

This paper is organized as follows. In Sec.~II, we derive expressions for the geometric phase of a two-level system undergoing pure dephasing for both initially pure and mixed system states. In Sec.~III, we compute the geometric phase for a two-level system interacting with an environment of harmonic oscillators both with and without initial system-environment correlations. A similar task is performed for a spin environment in Sec.~IV. Finally, we summarize our results in Sec.~V. Details regarding the exact solutions of the system-environment models employed are presented in the appendices. 

\section{The formalism}

\subsection{Pure initial system state}

Consider a two-level system with Hamiltonian $H_S$ interacting with an arbitrary environment whose Hamiltonian is $H_B$. The system-environment interaction is $H_{SB}$. The total system-environment Hamiltonian is then 
\begin{equation}
\label{totalHamiltonian}
H = H_S + H_B + H_{SB}. 
\end{equation}
For a pure dephasing model, $[H_S, H_{SB}] = 0$, which means that the in the eigenbasis of $H_S$, the diagonal elements of the density matrix of the two-level system do not change. In this basis, the initial state of the two-level system (assumed to be pure) can be written as  
\begin{align}
  \rho(0) =
  \left[ {\begin{array}{cc}
   \cos^2\left(\frac{\theta_0}{2}\right) & \frac{1}{2}\sin\theta_0 e^{-i\phi_0}  \\
    \frac{1}{2}\sin\theta_0 e^{i\phi_0}  &  \sin^2\left(\frac{\theta_0}{2}\right) \\
  \end{array} } \right].
  \label{pureinitialstate}
  \end{align}
Here $ 0\leq \theta_0 \leq \pi$, $ 0\leq \phi_0 <  2 \pi$ are the usual Bloch angles characterizing the initial system state. Since we are considering only pure dephasing, time evolution leads to a density matrix of the form 
\begin{align}
  \rho(t) =
  \left[ {\begin{array}{cc}
   \cos^2\left(\frac{\theta_0}{2}\right) & \frac{1}{2}\sin\theta_0 e^{-i\Omega(t)} e^{-\Gamma(t)} \\
    \frac{1}{2}\sin\theta_0 e^{i\Omega(t)} e^{-\Gamma(t)} &  \sin^2\left(\frac{\theta_0}{2}\right) \\
  \end{array} } \right].
  \label{finalstatewithpureinitialstate}
\end{align}
It is important to note that the density matrix $\rho(t)$ will have this form even in the presence of initial system-environment correlations - only the form of $\Omega(t)$ and $\Gamma(t)$ can be different. Now, in the Bloch vector representation, we can write $\rho(t)$ as  
$$ \rho(t) = \frac{1}{2}\left[ \mathds{1} + n_x \sigma_x + n_y \sigma_y + n_z \sigma_z \right], $$
where $n_x = \sin \theta_0 e^{-\Gamma(t)} \cos[\Omega(t)]$, $n_y = \sin \theta_0 e^{-\Gamma(t)} \sin[\Omega(t)]$, and $n_z = \cos \theta_0$. Given the density matrix $\rho(t)$, we can compute the geometric phase $\Phi_G$ via \cite{TongPRL2004} 
\begin{align}
\Phi_G &= \notag \\  
&\text{arg} \left(\sum_{k = 1}^2 \sqrt{\varepsilon_k(0) \varepsilon_k(\tau)} \ip{\varepsilon_k(0)}{\varepsilon_k(\tau)} e^{-\int_0^\tau dt\, \opav{\varepsilon_k}{\frac{\partial}{\partial t}}{\varepsilon_k}}\right). 
\end{align}
Here $\varepsilon_k(t)$ are the eigenvalues of the density matrix $\rho(t)$, $\ket{\varepsilon_k(t)}$ are the eigenvectors, and $\tau$ is the time after which the system completes a cyclic evolution. For our case, the eigenvalues of $\rho(t)$ are 
\begin{equation}
\varepsilon_\pm(t)  = \frac{1}{2}\bigg(1\pm \sqrt{1 + \sin^2 \theta_0 [e^{-2\Gamma(t)} - 1]}\bigg). 
\label{evpure}
\end{equation}
Notice that the eigenvalues are independent of $\Omega(t)$. Moreover, since $\varepsilon_-(0) = 0$, as is expected for a pure initial system state, our calculation of the geometric phase greatly simplifies. The corresponding eigenvectors of $\rho(t)$ are 
\begin{align}
\ket{\varepsilon_+(t)} &= \cos\left(\frac{\theta}{2}\right)\ket{0} + e^{i\Omega(t)}\sin\left(\frac{\theta}{2}\right)\ket{1}, \\
\ket{\varepsilon_-(t)} &= \sin\left(\frac{\theta}{2}\right)\ket{0} - e^{i\Omega(t)}\cos\left(\frac{\theta}{2}\right)\ket{1},
\end{align}
where
\begin{align*}
    \sin\theta&= \mathcal{F}(t)^{-1}\sin\theta_{0}e^{-\Gamma(t)},\\
    \cos\theta&=\mathcal{F}(t)^{-1}\cos\theta_{0},\\
    \mathcal{F}(t)&=\sqrt{1+\sin^{2}\theta_{0}(e^{-2\Gamma(t)}-1)},
\end{align*}
and $\ket{0}$ and $\ket{1}$ are the eigenstates of $H_S$. Since $\varepsilon_-(0) = 0$,  
$$ \Phi_G = \text{arg} \left(\sqrt{\varepsilon_+(0) \varepsilon_+(\tau)} \ip{\psi_+(0)}{\psi_+(\tau)} e^{-\int_0^\tau dt\, \opav{\psi_+}{\frac{\partial}{\partial t}}{\psi_+}}\right).$$
This further simplifies to 
$$ \Phi_G = \text{arg} \left(\ip{\varepsilon_+(0)}{\varepsilon_+(\tau)} e^{-\int_0^\tau dt\, \opav{\varepsilon_+}{\frac{\partial}{\partial t}}{\varepsilon_+}}\right),$$
since $\sqrt{\varepsilon_+(0) \varepsilon_+(\tau)}$ is real. We also find that $\opav{\varepsilon_+}{\frac{\partial}{\partial t}}{\varepsilon_+} = i\dot{\Omega} \sin^2 \left(\frac{\theta}{2}\right)$, where the dot denotes the time derivative. Moreover, 
\begin{align*}
 \ip{\varepsilon_+(0)}{\varepsilon_+(\tau)} &= \cos\left(\frac{\theta}{2}\right) \cos\left(\frac{\theta_0}{2}\right) \notag\\&+ e^{i\Omega(\tau)} e^{-i\phi_0} \sin\left(\frac{\theta}{2}\right)\sin\left(\frac{\theta_0}{2}\right).
 \end{align*}
The geometric phase can then be written as 
\begin{equation}
 \Phi_G = \Phi_1 + \Phi_2, 
 \label{GPprojective}
 \end{equation}
with $\Phi_1 = -\int_0^\tau dt \, \dot{\Omega} \sin^2 \left(\frac{\theta}{2}\right)$, and $\Phi_2 = \text{arg}\left[1 + e^{i\Omega(\tau)} e^{-i\phi_0} \tan\left(\frac{\theta}{2}\right) \tan\left(\frac{\theta_0}{2}\right)\right]$. To evaluate each of these one by one, we first note that $H_S$ has a characteristic frequency $\omega_0$ such that $\omega_0 \tau = 2\pi$. Then, $\Omega(t)$ can be written as $\Omega(t) = \phi_0 + \omega_0 t + \chi(t)$, where $\chi(t)$ takes into account part of the effect of the system-environment coupling. It follows that
\begin{align*}
\Phi_1 = -\int_0^\tau dt \,(\omega_0 +\dot{\chi}) \sin^2\left(\frac{\theta}{2}\right),
\end{align*}
which can be simplified to 
\begin{align}
\Phi_1 &= -\pi-\frac{\chi(\tau)}{2}+\frac{\cos\theta_{0}}{2}I(\tau),
\end{align}
with 
\begin{align*}
I(\tau)&=\int_{0}^{\tau}\mathcal{F}(t)^{-1}[\omega_{0}+\dot{\chi}(t)]\>dt.
\end{align*}
As for $\Phi_2$, we can write
\begin{align*}
    \Phi_{2}=\text{arg}\bigg(1+e^{i\chi(\tau)}\tan\bigg[ \frac{\theta(0)}{2}\bigg]\tan\bigg[ \frac{\theta(\tau)}{2}\bigg] \bigg).
\end{align*}
Since $\tan \left(\frac{\theta}{2}\right) = \frac{\sin \theta}{1 + \cos \theta}$ and $\tan \theta = (\tan \theta_0) e^{-\Gamma(t)}$, this further simplifies to 
\begin{align}
\Phi_2& = \text{arg}\bigg(1+e^{i\chi(\tau)}e^{-\Gamma(\tau)}\frac{1-\cos\theta_{0}}{\mathcal{F}(\tau)+\cos\theta_{0}}\bigg). 
\end{align}
With $\Phi_1$ and $\Phi_2$ found, we can thereby calculate $\Phi_G$. It should be noted that if the system-environment interaction strength is zero, we find that $\Phi_1 = -\pi + \pi \cos \theta_0$ while $\Phi_2 = 0$, thereby leading to the usual result $\Phi_G = -\pi + \pi \cos \theta_0$ \cite{AharonovPRL1987}. Moreover, for $\theta_0 = \pi/2$, $\Phi_{1}= -\pi-\frac{\chi(\tau)}{2}$ and $\Phi_2 = \frac{\chi(\tau)}{2}$, meaning that $\Phi_G = -\pi$. Thus the geometric phase is robust for the states with $\theta_0 = \pi/2$ even if initial correlations are taken into account. Consequently, we will consider $\theta_0 \neq \pi/2$ to investigate the effect of the initial correlations on the geometric phase. Before doing so for concrete system-environment models, we generalize our results to the case where the initial state is mixed. 

\subsection{Mixed initial system state}
We now derive expressions for the geometric phase for initially mixed states. Our approach will be to write the state for the two-level system in a form similar to that in Eqs.~\eqref{pureinitialstate} and \eqref{finalstatewithpureinitialstate} so that we obtain an expression for the geometric phase similar to that in Eq.~\eqref{GPprojective}. As such, we start by noting that the initial density matrix, even for a mixed state, can be written as 
\begin{align}
\rho(0) = 
   \begin{pmatrix}
    \cos^{2}\bigg(\dfrac{\Tilde{\theta}_{0}}{2} \bigg)& \frac{1}{2}e^{-\Gamma_{0}}\sin\Tilde{\theta}_{0}e^{-i\phi_{0}}\\
    \frac{1}{2}e^{-\Gamma_{0}}\sin\Tilde{\theta}_{0}e^{i\phi_{0}}& \sin^{2}\bigg(\dfrac{\Tilde{\theta}_{0}}{2} \bigg).
  \end{pmatrix}.
\end{align}
Note that $\tilde{\theta}_0$ is not a Bloch angle here. $\Gamma_0 > 0$ takes into account that the initial state is mixed. It follows that 
\begin{align*}
\rho(t) = 
   \begin{pmatrix}
    \cos^{2}\bigg(\dfrac{\Tilde{\theta}_{0}}{2} \bigg)& \frac{1}{2}\sin\Tilde{\theta}_{0}e^{-i\Omega(t)-\Gamma_{0}-\Gamma(t)}\\
    \frac{1}{2}\sin\Tilde{\theta}_{0}e^{i\Omega(t) -\Gamma_{0} - \Gamma(t)} & \sin^{2}\bigg(\dfrac{\Tilde{\theta}_{0}}{2} \bigg)\\
  \end{pmatrix},
\end{align*}
with $\Omega(t) = \omega_{0}t+\chi(t)+\phi_{0}$ as before. The eigenvalues of the density matrix $\rho(t)$ are now a simple extension of Eq.~\eqref{evpure}, that is,
\begin{equation}
\varepsilon_{\pm}=\frac{1}{2}\bigg(1\pm \Tilde{\mathcal{F}}(t) \bigg),
\end{equation}
with 
$$\Tilde{\mathcal{F}}(t)=\sqrt{1+\sin^{2}\Tilde{\theta}_{0}(e^{-2\Gamma_{0}}e^{-2\Gamma(t)}-1)}.$$
The corresponding eigenvectors are similarly 
\begin{align*}
    \ket{\varepsilon_{+}} &=\cos\bigg(\frac{\Tilde{\theta}}{2}\bigg)\ket{0}+e^{i\Omega(t)}\sin\bigg(\frac{\Tilde{\theta}}{2}\bigg)\ket{1}, \\
    \ket{\varepsilon_{-}} &=\sin\bigg(\frac{\Tilde{\theta}}{2}\bigg)\ket{0} - e^{i\Omega(t)}\cos\bigg(\frac{\Tilde{\theta}}{2}\bigg)\ket{1}, \\
\end{align*}
where $\sin\Tilde{\theta}=\sin\Tilde{\theta}_{0}e^{-\Gamma_{0}}e^{-\Gamma(t)}{\Tilde{\mathcal{F}}(t)}^{-1}$ and $\cos\Tilde{\theta}=\cos\Tilde{\theta}_{0}{\Tilde{\mathcal{F}}(t)}^{-1}$. With the density matrix $\rho(t)$ found, the geometric phase $\Phi_G$ can be written as 
\begin{equation}
\Phi_G = \Phi_1 + \Phi_2 + \Phi_3,
\end{equation}
with 
\begin{align*}
\Phi_1 &=\arg \bigg(e^{-\int_0^\tau dt\, \opav{\varepsilon_+}{\frac{\partial}{\partial t}}{\varepsilon_+}}\bigg), \\
\Phi_2 &=\arg \langle \varepsilon_{+}(0)|\varepsilon_{+}(\tau)\rangle, \\
 \Phi_{3}&=\arg\bigg( 1+\sqrt{\frac{\varepsilon_-(0) \varepsilon_-(\tau)}{\varepsilon_+(0) \varepsilon_+(\tau)}}\frac{\ip{\varepsilon_-(0)}{\varepsilon_-(\tau)}}{\ip{\varepsilon_+(0)}{\varepsilon_+(\tau)}} \, \times \, \\
 &e^{\int_0^\tau dt\, \opav{\varepsilon_+}{\frac{\partial}{\partial t}}{\varepsilon_+}-\opav{\varepsilon_-}{\frac{\partial}{\partial t}}{\varepsilon_-}} \bigg). \label{three}
\end{align*}
The calculations for $\Phi_1$ and $\Phi_2$ can be performed as done before to obtain 
\begin{align}
\Phi_{1}&=-\pi-\frac{\chi(\tau)}{2}+\frac{1}{2}\cos\Tilde{\theta}_{0}\Tilde{I}(\tau),
\end{align}
where 
$$\Tilde{I}(\tau) =  \int_{0}^{\tau}\, dt \, \frac{\omega_{0}+\dot{\chi}}{\sqrt{1+\sin^{2}\Tilde{\theta}_{0}(e^{-2\Gamma_{0}}e^{-2\Gamma(t)}-1)}}, $$
and 
\begin{align}
    \Phi_{2}=\arg\Bigg(1 +e^{i\chi(\tau)}\tan\bigg[\frac{\Tilde{\theta}(0)}{2}\bigg]\tan\bigg[\frac{\Tilde{\theta}(\tau)}{2}\bigg]   \Bigg).
\end{align}
Finally, we compute $\Phi_{3}$ and find that 
\begin{align}
    \Phi_{3}&=\arg\bigg( 1+ a(\tau) b(\tau) e^{-i\cos\Tilde{\theta}_{0}\Tilde{I}(\tau)} \bigg),
\end{align}
where
$$ a(\tau) = \sqrt{\frac{\varepsilon_-(0) \varepsilon_-(\tau)}{\varepsilon_+(0) \varepsilon_+(\tau)}}, $$
and 
$$ b(\tau) =\frac{\tan\bigg[\dfrac{\Tilde{\theta}(0)}{2}\bigg]\tan\bigg[\dfrac{\Tilde{\theta}(\tau)}{2}\bigg]+e^{i\chi(\tau)}}{1+e^{i\chi(\tau)}\tan\bigg[\dfrac{\Tilde{\theta}(0)}{2}\bigg]\tan\bigg[\dfrac{\Tilde{\theta}(\tau)}{2}\bigg]}. $$
Finding the geometric phase now is simply a matter of finding the parameters $\tilde{\theta}_0$, $\phi_0$, and $\Gamma_0$ characterizing the initial state as well as the functions $\Gamma(t)$ and $\chi(t)$ that go into the time evolution of the system density matrix. It is important to realize that if the system-environment interaction is zero, the geometric phase is, in general, no longer $-\pi + \pi \cos \Tilde{\theta}_0$ since the initial state is mixed. However, for $\tilde{\theta}_0 = \pi/2$, we again obtain $\Phi_G = -\pi$.

We will now use the expressions for the geometric phase to perform calculations with concrete system-environment models, both with and without initial system-environment correlations.

\section{Two-level system interacting with an environment of harmonic oscillators}

We first apply our formalism to the paradigmatic example of a single two-level system undergoing pure dephasing via interaction with a collection of harmonic oscillators \cite{BPbook}. The total system-environment Hamiltonian is $H = H_S + H_B + H_{SB}$, where (we set $\hbar = 1$ throughout) 
\begin{align*} 
 &H_S = \frac{\omega_0}{2} \sigma_z, \; H_B = \sum_k \omega_k b_k^\dagger b_k, \\
 &H_{SB} = \sigma_z  \sum_k (g_k^* b_k + g_k b_k^\dagger), 
\end{align*}
and $\sigma_z$ is the usual Pauli matrix, $\omega_0$ is the energy bias, and $b_k$ ($b_k^\dagger$) are the annihilation (creation) operators for the harmonic oscillator modes. Since $[H_S,H_{SB}] = 0$, $\langle \sigma_z \rangle$ does not change with time, and only dephasing takes place. Assuming that the initial system-environment state is a product state with the environment in a thermal equilibrium state $\rho_B = e^{-\beta H_B}/Z_B$, where $Z_B = \tr_B [e^{-\beta H_B}]$, the evolution of the off-diagonal elements of the density matrix is given by \cite{BPbook}  
\begin{equation}
\langle \sigma_{\pm}(t)\rangle=\langle \sigma_{\pm}\rangle e^{\pm i\omega_{0}t}e^{-\Gamma_{\text{uc}}(t)}, 
\label{hoenvuc}
\end{equation}
where 
$$ \Gamma_{\text{uc}}(t)= \sum_k 4 |g_k|^2 \coth(\beta \omega_k/2)\frac{1-\cos\omega_k t}{\omega_k^{2}}. $$
For completeness, the derivation of this result is presented in Appendix \ref{appendixhoenv}. On the other hand, if the system and the environment have interacted for a long time beforehand, the initial state of the environment is not the thermal equilibrium state $e^{-\beta H_B}/Z_B$. Instead, the system and the environment together are in a thermal equilibrium state, that is, $e^{-\beta H}/Z$, where $Z = \tr_{S,B} [e^{-\beta H}]$ \cite{Weissbook}. Then, at time $t = 0$, we can perform either a projective measurement or a unitary operation on the system to prepare the desired initial system state. We now analyze these scenarios one by one. 

\subsection{System state preparation by projective measurement}

\begin{figure}[t!]
\begin{minipage}[t]{0.48\linewidth}
\includegraphics[width=.9\linewidth]{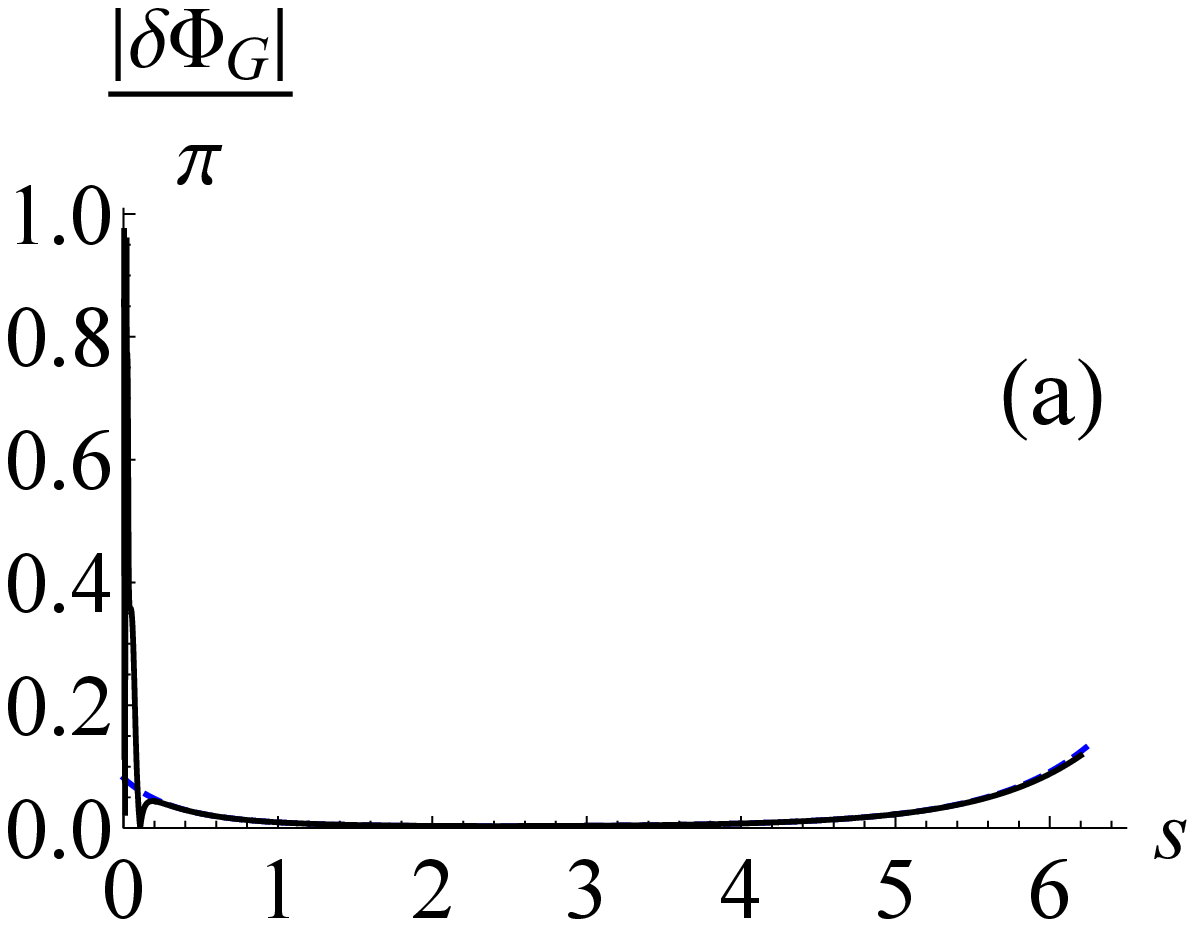}
\end{minipage}\hfill%
\begin{minipage}[t]{0.48\linewidth}
\includegraphics[width=.9\linewidth]{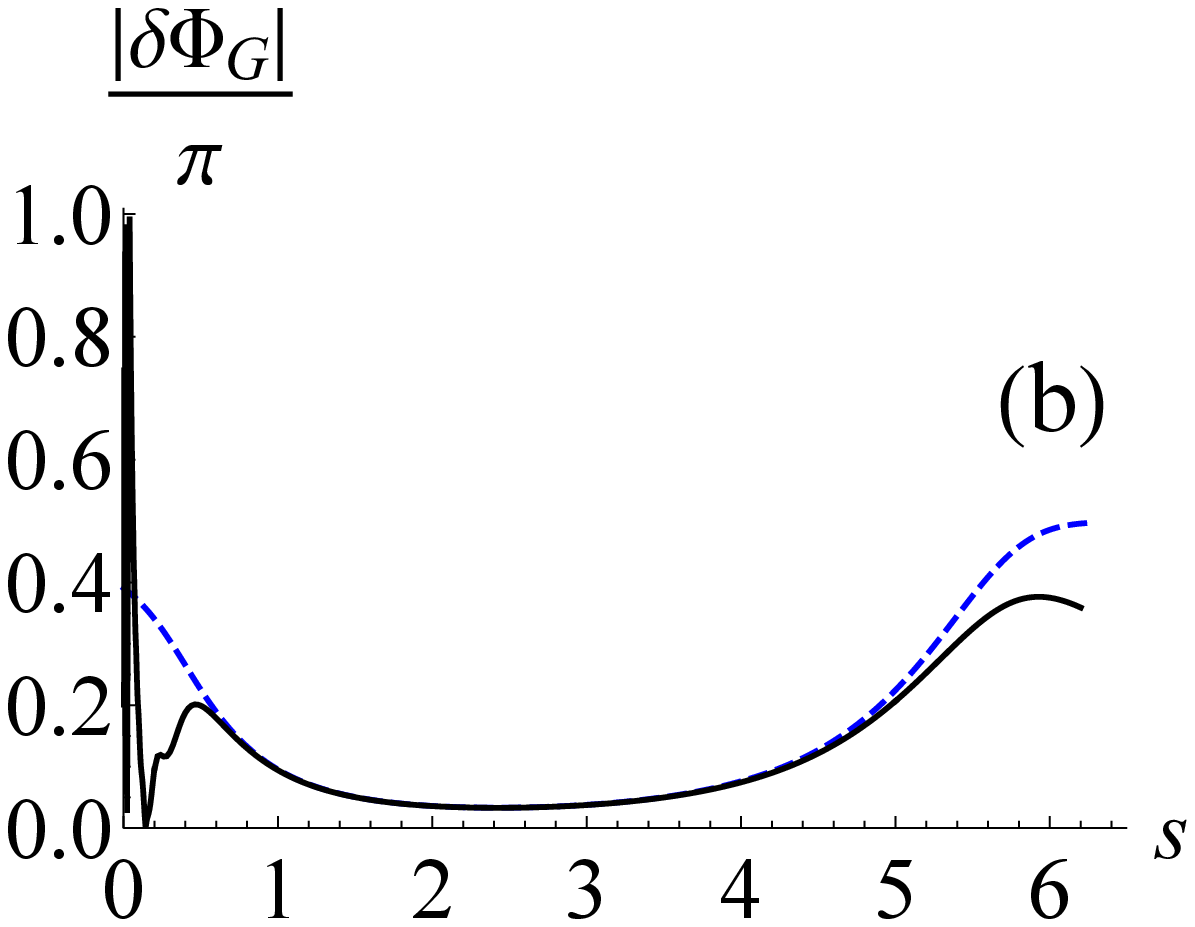}
\end{minipage}%
\caption{(Color online) Correction to the geometric phase $\delta \Phi_G \equiv \Phi_G - \Phi_U$ (where $\Phi_U = -\pi + \pi \cos \theta_0$) of the two-level system in the presence of the harmonic oscillator environment as a function of the Ohmicity parameter $s$ for weak system-environment coupling strength. The solid, black curve shows the geometric phase when the initial state is prepared via a projective measurement, while the dashed, blue curve is for an uncorrelated initial state [that is, the dynamics are given by Eq.~\eqref{hoenvuc}]. In (a), the system-environment coupling strength is $\lambda=0.01$, while in (b) we have used $\lambda=0.1$. Throughout, we are working in dimensionless units with $\hbar = 1$, and here we have set $\omega_0 = 1$. We have used $\omega_c=5$, $\theta_0=\pi/3$, and $\beta \rightarrow \infty$ (zero temperature).}
\label{fighoenvzerotempweak}
\end{figure}

\begin{figure}[t!]
\begin{minipage}[t]{0.48\linewidth}
\includegraphics[width=.9\linewidth]{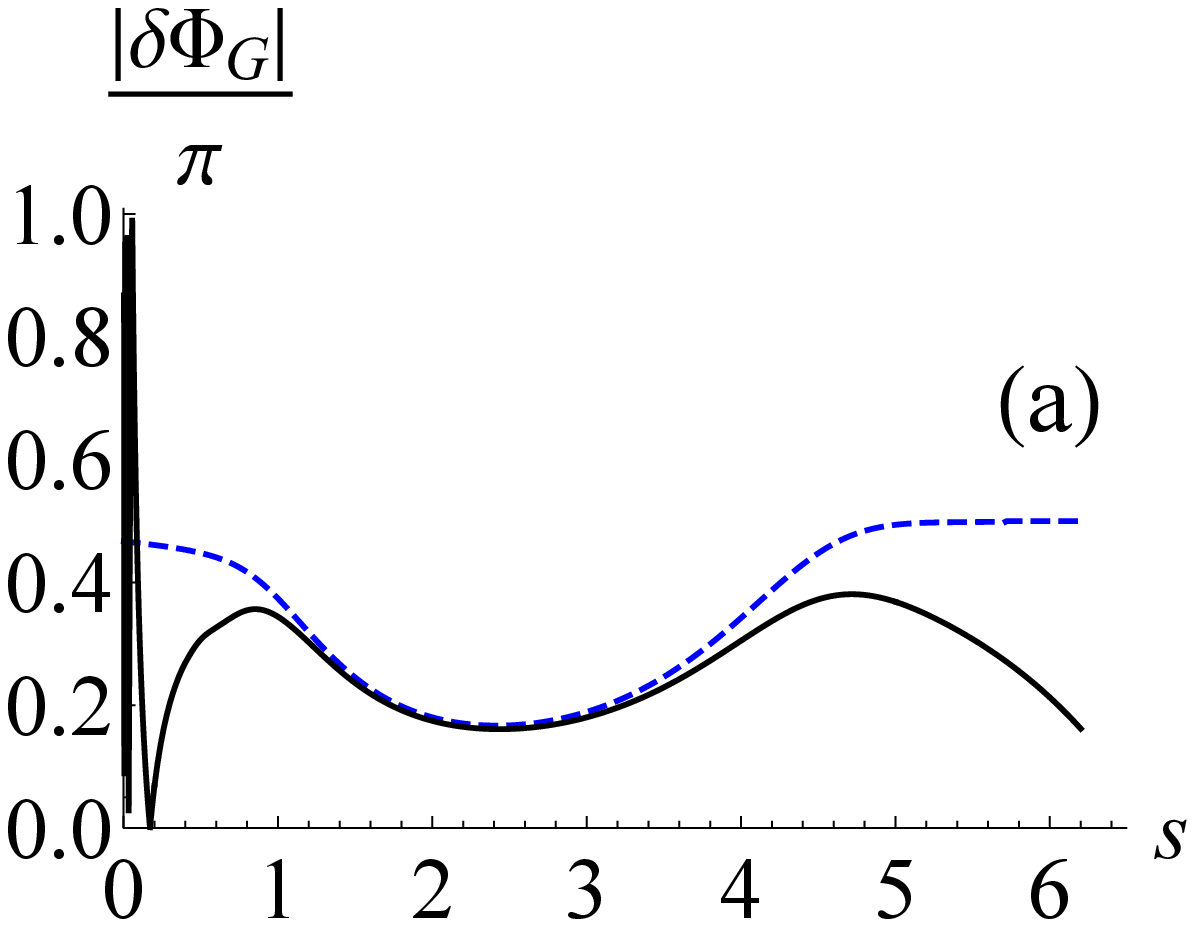}
\end{minipage}\hfill%
\begin{minipage}[t]{0.48\linewidth}
\includegraphics[width=.9\linewidth]{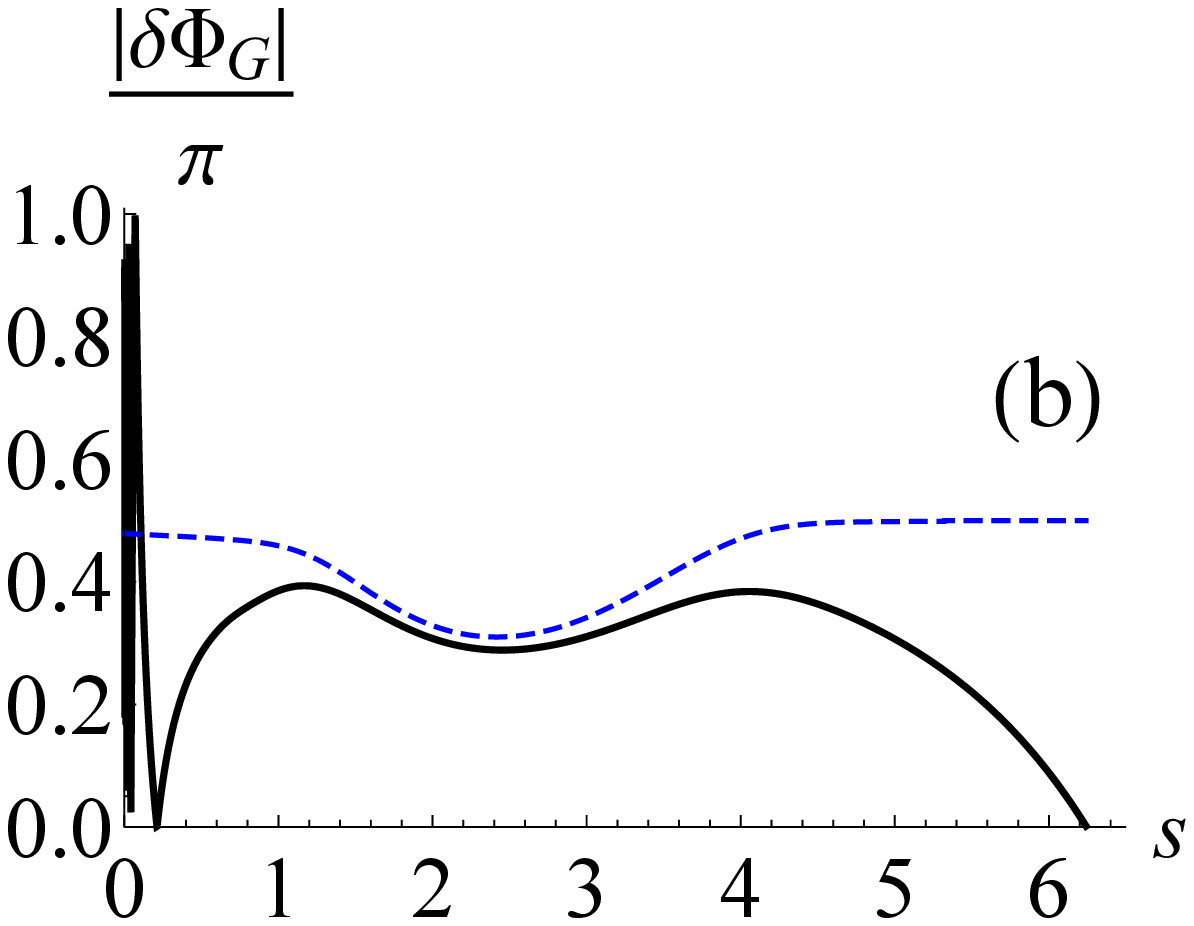}
\end{minipage}%
\caption{(Color online) Same as Fig \ref{fighoenvzerotempweak} except that in (a), we have $\lambda=0.5$ and in (b), we have $\lambda=1$.}
\label{fighoenvzerotempmodstrong}
\end{figure}

\begin{figure}[t!]
\begin{minipage}[t]{0.48\linewidth}
\includegraphics[width=.9\linewidth]{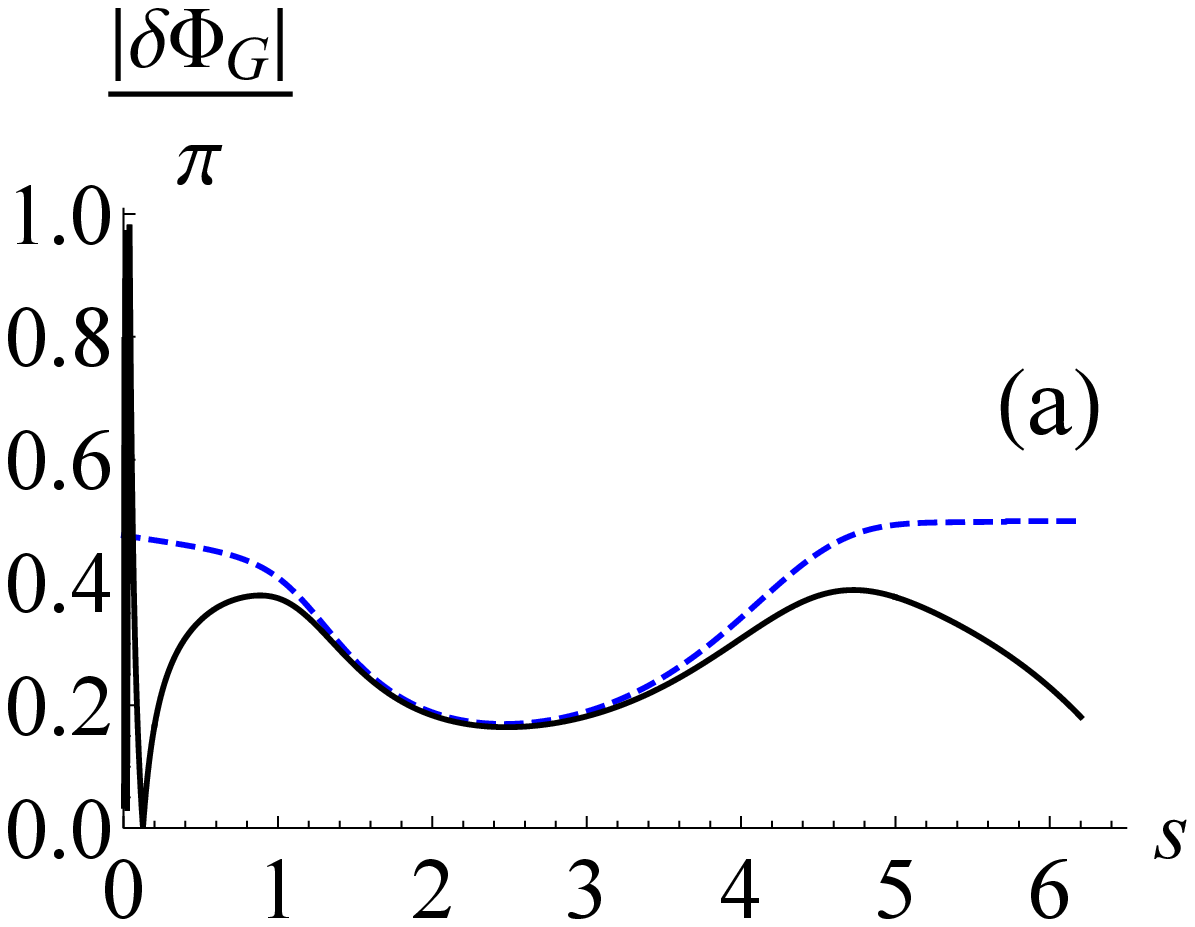}
\end{minipage}\hfill%
\begin{minipage}[t]{0.48\linewidth}
\includegraphics[width=.9\linewidth]{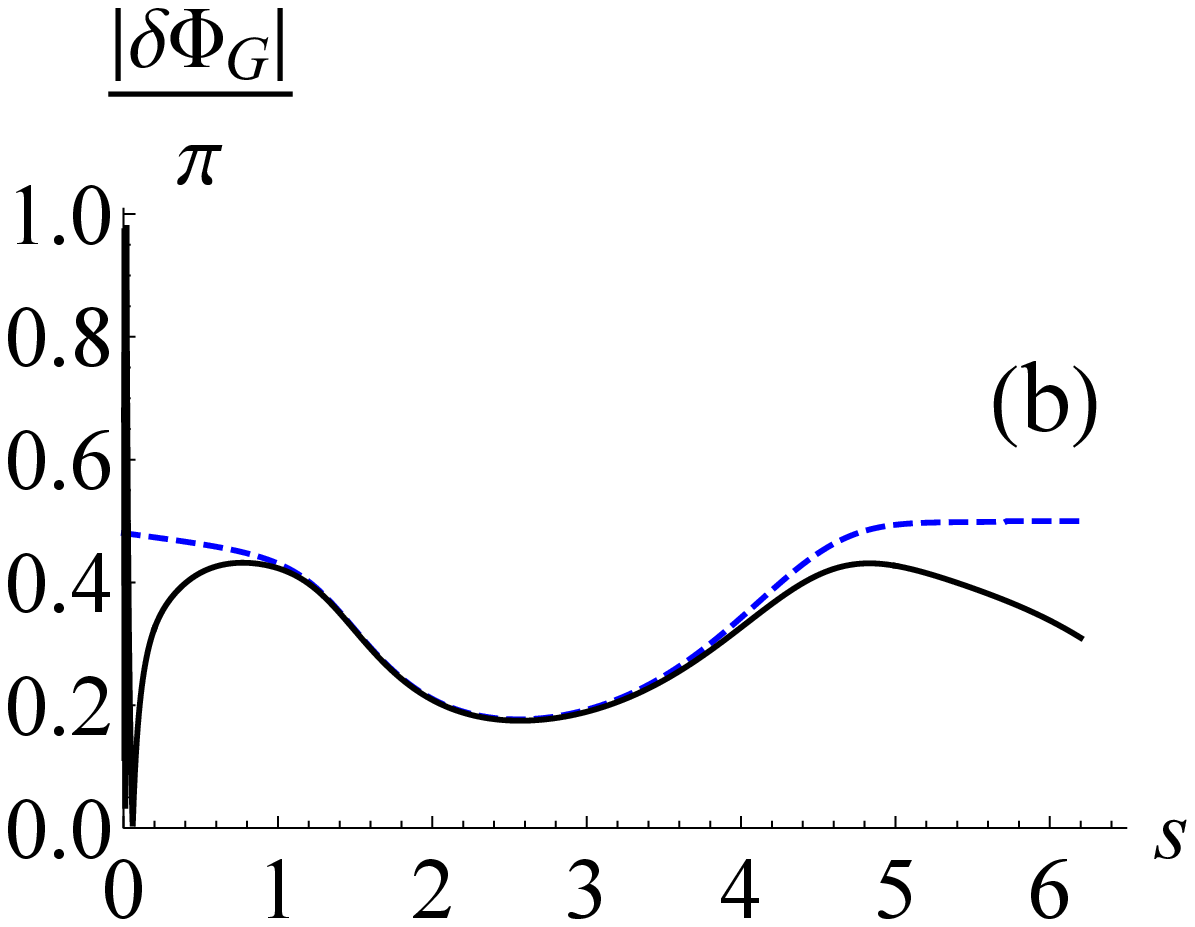}
\end{minipage}%
\caption{(Color online) Same as Fig.~\ref{fighoenvzerotempweak}, except that the system-environment coupling strength is $\lambda=0.5$, and in (a), we have $\beta=2$, while in (b) we have $\beta=1$.}
\label{fighoenvnonzerotempmod}
\end{figure}

\begin{figure}[h!]
\begin{minipage}[t]{0.48\linewidth}
\includegraphics[width=\linewidth]{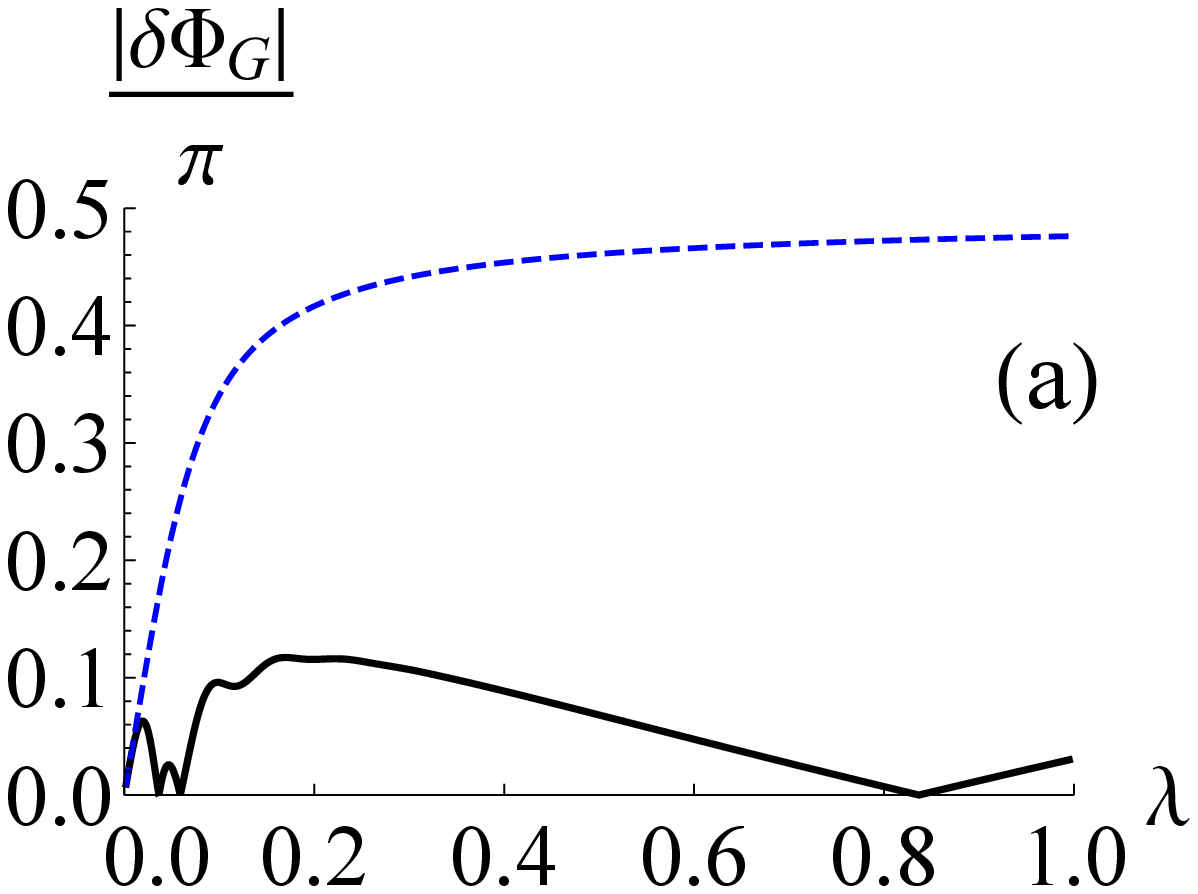}
\end{minipage}\hfill%
\begin{minipage}[t]{0.48\linewidth}
\includegraphics[width=\linewidth]{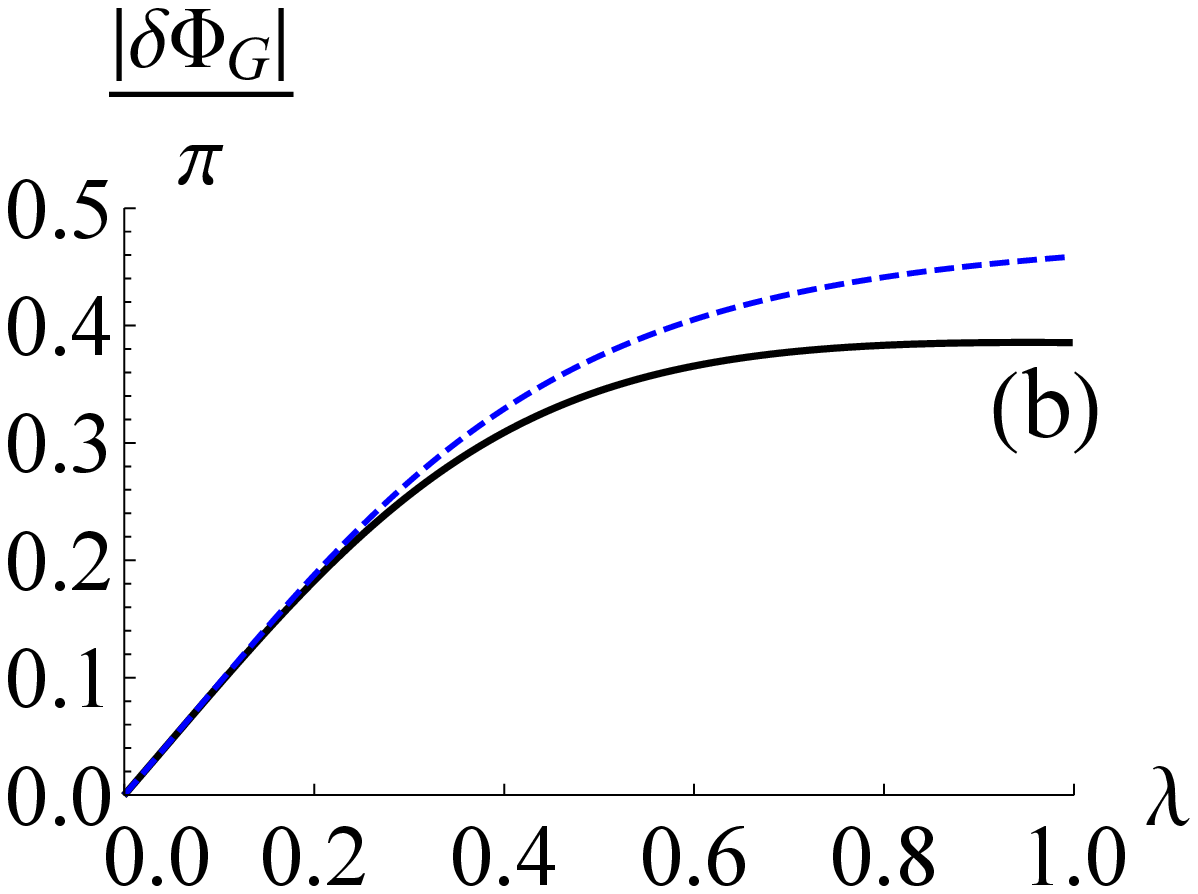}
\end{minipage}%
\caption{Color online) Correction to the geometric phase $\delta \Phi_G \equiv \Phi_G - \Phi_U$ (where $\Phi_U = -\pi + \pi \cos \theta_0$) with a harmonic oscillator environment as a function of the system-environment coupling strength $\lambda$. The solid, black curve shows the geometric phase when the initial state is prepared via a projective measurement, while the dashed, blue curve is for an uncorrelated initial state. In (a), the Ohmicity parameters is $s = 0.2$, while for (b), $s = 1$. These results are obtained for $\beta \rightarrow \infty$ (zero temperature). As before, we have set $\omega_0 = 1$, and we have used $\omega_c=5$ and $\theta_0=\pi/3$.}
\label{fighoenvvaryingsysenvcoupling}
\end{figure}

If the initial system state $\ket{\psi}$ is prepared by a projective measurement, described by the projector $P_\psi = \ket{\psi}\bra{\psi}$, then the initial system-environment state is $\rho(0) = \frac{1}{Z}P_\psi e^{-\beta H}P_\psi$ with $Z= \tr_{S,B} [P_\psi e^{-\beta H}]$. With this initial state, the evolution of the off-diagonal elements of the system density matrix is given by \cite{MorozovPRA2012,ChaudhryPRA2013a} 
\begin{align}
\label{hoenvprojectiveprep}
    \langle \sigma_{\pm}(t)\rangle=\langle \sigma_{\pm}\rangle e^{\pm i[\omega_{0}t+\chi(t)]}e^{-\Gamma(t)},
\end{align}
where 
$$    \Gamma(t)=\Gamma_{\text{uc}}(t)+\Gamma_{\text{corr}}(t), $$
\begin{align*}
    &\Gamma_{\text{corr}}(t)= \\
    &-\frac{1}{2}\ln\bigg[1-\frac{(1-\cos^2\theta_{0})\sin^2[\Phi(t)]}{[\cosh(\beta \omega_{0}/2)-\cos\theta_{0}\sinh(\beta\omega_{0}/2)]^2} \bigg],
\end{align*}
\begin{align*}
    \tan[\chi(t)]&=\frac{\sinh(\beta \omega_{0}/2)-\cos\theta_{0}\cosh(\beta\omega_{0}/2)}{\cosh(\beta \omega_{0}/2)-\cos\theta_{0}\sinh(\beta \omega_{0}/2)}\tan[\Phi(t)], \\
    \Phi(t)&=\sum_{k}\frac{4|g_{k}|^2}{\omega_{k}^2}\sin(\omega_kt).
\end{align*}
For completeness, the derivation of these results is sketched in Appendix \ref{appendixhoenv}. Note that the effect of the initial correlations is to modify the decoherence rate as well as to introduce a phase shift. Moreover, for zero temperature, these expressions further simplify to $\Gamma_{\text{corr}}(t) = 0$ and $\chi(t) = \Phi(t)$. In this case, the decoherence rate is not modified and the effect of the initial correlations is a simple phase shift given by $\Phi(t)$.

With the system density matrix found, the geometric phase can then be evaluated. To calculate the sum over the environment modes, the sum is converted to an integral via the spectral density $J(\omega)$, which allows us to write $\sum_k 4|g_k|^2 (\hdots)$ as $\int_0^\infty \, d\omega \, J(\omega) (\hdots)$. We consider the spectral density to be of the form $J(\omega) = \lambda\omega^s \omega_c^{1-s} e^{-\omega/\omega_c}$, where $\lambda$ is a dimensionless constant characterizing the system-environment interaction strength, $s$ is the so-called Ohmicity parameter, and $\omega_c$ is the cutoff frequency \cite{BPbook}. In Figs.~\ref{fighoenvzerotempweak}(a) and (b), we have plotted the behavior of the correction to the geometric phase $|\delta \Phi_G|$, as the Ohmicity parameter is varied, for weak system-environment coupling strength. It is clear from these figures that for weak system-environment coupling strength, the effect of the initial correlations on the geometric phase is generally negligible since the dashed blue line largely overlaps with the solid black curve. Nevertheless, for sub-Ohmic environments (that is, $s < 1$), the initial correlations can still play a role. Interestingly, taking the initial correlations into account generally makes the correction to the geometric phase smaller. In fact, for a particular value of the Ohmicity parameter, the correction to the geometric phase is zero. Proceeding along these lines, in Figs.~\ref{fighoenvzerotempmodstrong}(a) and (b) we have shown the correction to the geometric phase at zero temperature for stronger system-environment coupling strengths. Three points are evident from these figures. First, for a range of values of $s$, the initial correlations have a very small effect on the geometric phase. Second, for sub-Ohmic environments as well as for very super-Ohmic environments, the contribution of the initial correlations to the geometric phase is very significant. Third, the initial correlations generally reduce the correction to the geometric phase, thereby implying that the initial correlations increase the robustness of the geometric phase. As before, for a particular value of the Ohmicity parameter, the correction to the geometric phase becomes zero. We have also found that, as expected, as the temperature is increased, the effect of the initial correlations decreases [see Figs.~\ref{fighoenvnonzerotempmod}(a) and (b)].

It is also interesting to analyze the correction to the geometric phase as the system-environment couping strength is varied. The results are illustrated in Fig.~\ref{fighoenvvaryingsysenvcoupling}(a) and (b). For the sub-Ohmic environment considered in Fig.~\ref{fighoenvvaryingsysenvcoupling}(a), the initial correlations greatly reduce the correction to the geometric phase. Surprisingly, as the system-environment coupling strength is increased, the correction to the geometric phase, in the case where the initial correlations are taken into account, can decrease. In fact, for particular non-zero values of the system-environment interaction strength, the correction to the geometric phase becomes zero. This is not the case for an Ohmic environment [see Fig.~\ref{fighoenvvaryingsysenvcoupling}(b)].

\subsection{System state preparation by unitary operation}

\begin{figure}
\begin{minipage}[t]{0.48\linewidth}
\includegraphics[width=.9\linewidth]{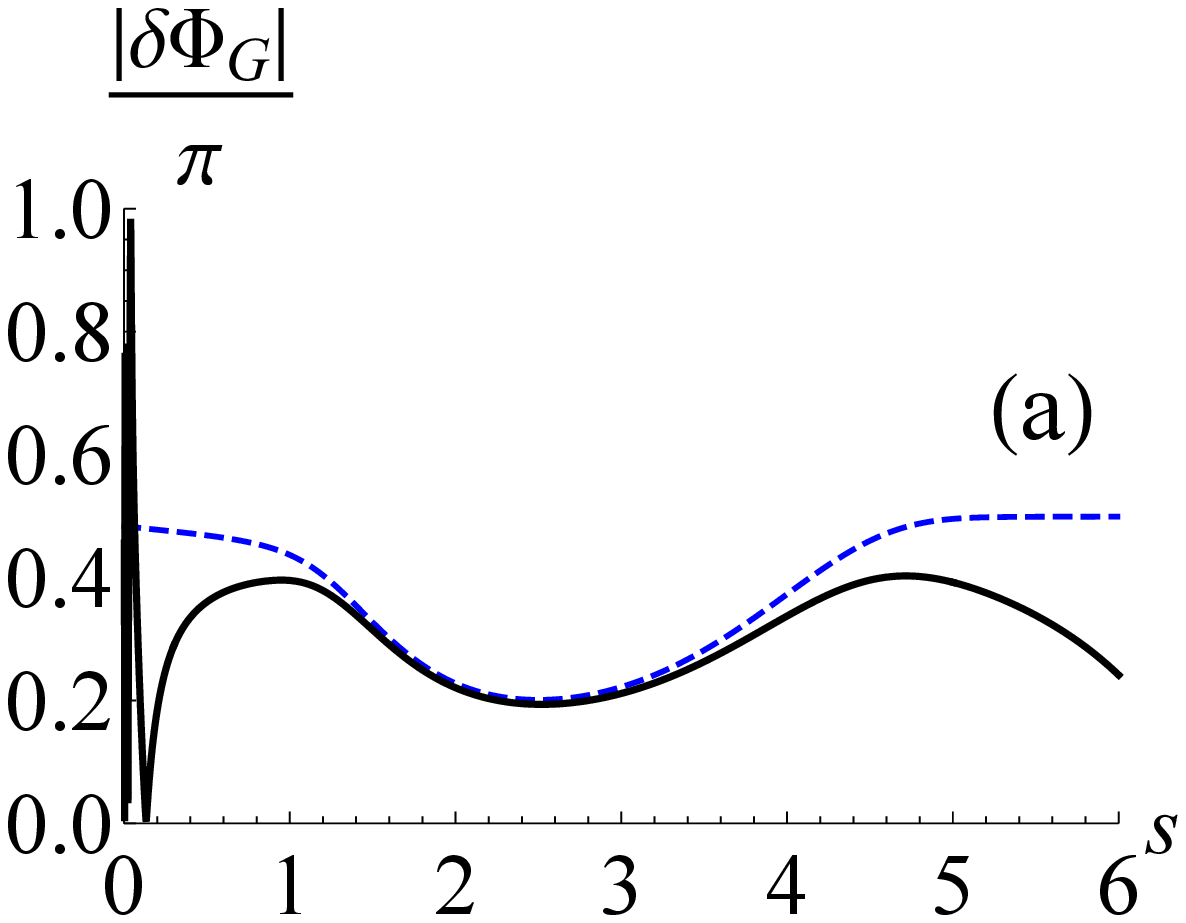}
\end{minipage}\hfill%
\begin{minipage}[t]{0.48\linewidth}
\includegraphics[width=.9\linewidth]{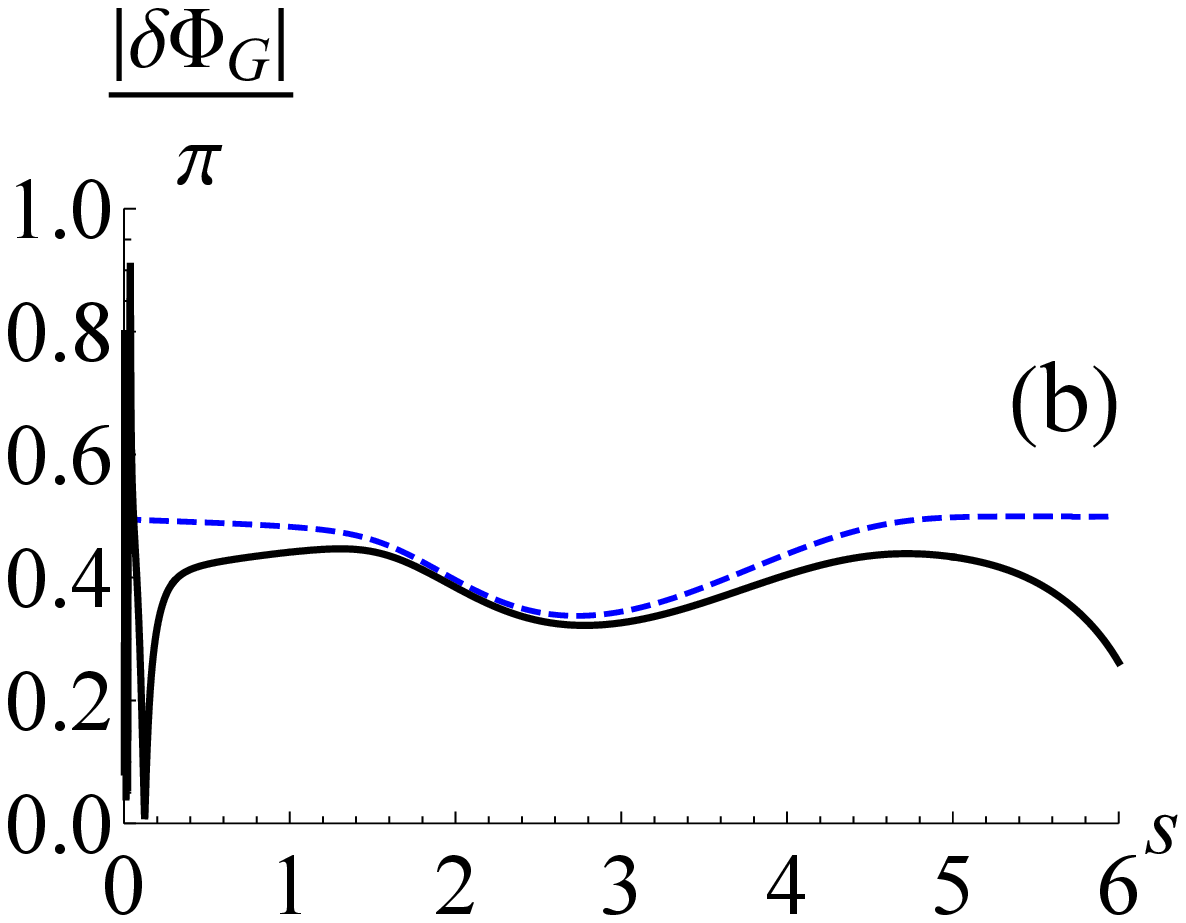}
\end{minipage}%
\caption{(Color online) Correction to the geometric phase $\delta \Phi_G \equiv \Phi_G - \Phi_0$ (where $\Phi_0$ is the geometric phase for the two-level system if the system-environment interaction strength is zero) with a harmonic oscillator environment as a function of the Ohmicity parameter $s$ if the initial state is prepared via a unitary operation. The solid, black curve shows the geometric phase when the initial state is prepared via the unitary operation $\Omega = e^{i \pi \sigma_y/3}$, while the dashed, blue curve is for an uncorrelated initial state. In (a), we have $\beta=3$ while in (b), we have $\beta=1$. Once again, we have set $\omega_0 = 1$, and we have used $\omega_c=5$ and $\lambda = 0.5$. }
\label{fighoenvunitaryprep}
\end{figure}

We now analyze the effect of the initial correlations if a unitary operation, instead of a projective measurement, is used to prepare the initial system state. The initial system-environment state in this case is $\rho(0) = \frac{1}{Z}\Omega e^{-\beta H} \Omega^\dagger$, where $\Omega$ is a unitary operation performed on the system. The off-diagonal elements of the system density matrix are given by \cite{MorozovPRA2012,ChaudhryPRA2013a}
\begin{equation}
\label{hoenvunitaryprep}
\langle \sigma_\pm (t) \rangle = \langle \sigma_\pm (0) \rangle e^{\pm i [\omega_0 t + \chi(t)]} e^{-\Gamma(t)},
\end{equation}
with
\begin{align}
    \Gamma(t)&=\Gamma_{\text{uc}}(t)+\Gamma_{\text{corr}}(t),
\end{align}
where
\begin{widetext}
\begin{align}
    \Gamma_{\text{corr}}(t)&=-\ln \left\lbrace\text{abs}\left[\frac{e^{-\beta \omega_0/2}  \opav{0}{\Omega^\dagger \sigma_+ \Omega}{0} e^{- i\Phi(t)} + e^{\beta \omega_0/2}  \opav{1}{\Omega^\dagger \sigma_+ \Omega}{1} e^{+ i\Phi(t)}}{e^{-\beta \omega_0/2}  \opav{0}{\Omega^\dagger \sigma_+ \Omega}{0}  + e^{\beta \omega_0/2}  \opav{1}{\Omega^\dagger \sigma_+ \Omega}{1}}\right]\right\rbrace, \label{gammacorrhounit}\\
    \chi(t) &= \arg \bigg[\cos[\Phi(t)]+i\sin[\Phi(t)] \bigg( \frac{\langle 1|\Omega^\dagger \sigma_+ \Omega|1\rangle e^{\beta \omega_0/2}-\langle 0|\Omega ^\dagger \sigma_+ \Omega|0\rangle e^{-\beta \omega_0/2}}{\langle 1|\Omega^\dagger \sigma_+ \Omega|1\rangle e^{\beta \omega_0/2}+\langle 0|\Omega ^\dagger \sigma_+ \Omega|0\rangle e^{-\beta \omega_0/2}}\bigg) \bigg] \label{phicorrhounit}.
\end{align}
\end{widetext}
Here $\ket{0}$ and $\ket{1}$ are the eigenstates of $\sigma_z$ with $\sigma_z \ket{l} = (-1)^l \ket{l}$. These derivations are again sketched in Appendix \ref{appendixhoenv}. One can check from these expressions that for zero temperature, $\Gamma_{\text{corr}} = 0$ and $\chi(t) = \Phi(t)$. Consequently, the behavior of the geometric phase at zero temperature is the same as when the initial system state is prepared via a projective measurement. However, there will be differences at non-zero temperatures. The correction to the geometric phase $\delta \Phi_G = \Phi_G - \Phi_0$, where $\Phi_0$ is the geometric phase for the two-level system if the system-environment coupling strength is zero, is plotted as a function of the Ohmicity parameter $s$ for two different temperatures in Fig.~\ref{fighoenvunitaryprep} for moderate system-environment coupling strength. Once again, it is clear that the initial correlations can play a very significant role for the geometric phase, especially for sub-Ohmic environments. 

\section{Two-level system interacting with spin environment}

We now consider the central two-level system to be interacting with a collection of $N$ two-level systems \cite{cucchietti2005decoherence,CamaletPRB2007, Schlosshauerbook, VillarPhysLettA2009}. The system Hamiltonian $H_S$ is still $\frac{\omega_0}{2}\sigma_z$, while the environment Hamiltonian is now $\sum_{i}\omega_{i}\sigma^{i}_{x}$, and the system-environment interaction is described by $\sigma_{z}\sum_{i}\lambda_{i}\sigma_{z}^{i}$. Since $[H_S, H_{SB}] = 0$, this is also a pure dephasing model. If the initial system-environment state is a product state of the form $\rho(0) = \rho_S(0) \otimes e^{-\beta H_B}/Z_B$, then the evolution of the off-diagonal elements is given by \cite{CamaletPRB2007,VillarPhysLettA2009}
\begin{align*}
    \langle \sigma_{\pm}(t)\rangle&=\langle \sigma_{\pm}\rangle e^{\pm i\omega_{0}t}e^{-\Gamma_{\text{uc}}(t)},
\end{align*}
where 
\begin{equation*}
    \Gamma_{\text{uc}}(t)=-\sum_{j}\ln\bigg\{1- \frac{2\lambda_{j}^2}{\lambda_{j}^2+\omega_{j}^2}\sin^2(\sqrt{\lambda_{j}^2+\omega_{j}^2}t) \bigg\},
\end{equation*}
and the sum is over the environment spins. The derivation of this result is reproduced in Appendix \ref{appendixspinenv}. However, as emphasized before, this result may questionable since the initial system-environment correlations are disregarded. To investigate the effect of these correlations, we consider the system state to be prepared by a projective measurement as well as by a unitary operation starting from the total system-environment equilibrium state $e^{-\beta H}/Z$. We note that, to the best of our knowledge, this model has not been solved taking initial correlations into account before.

\subsection{System state preparation by projective measurement}

\begin{figure}[b!]
\begin{minipage}[t]{0.48\linewidth}
\includegraphics[width=\linewidth]{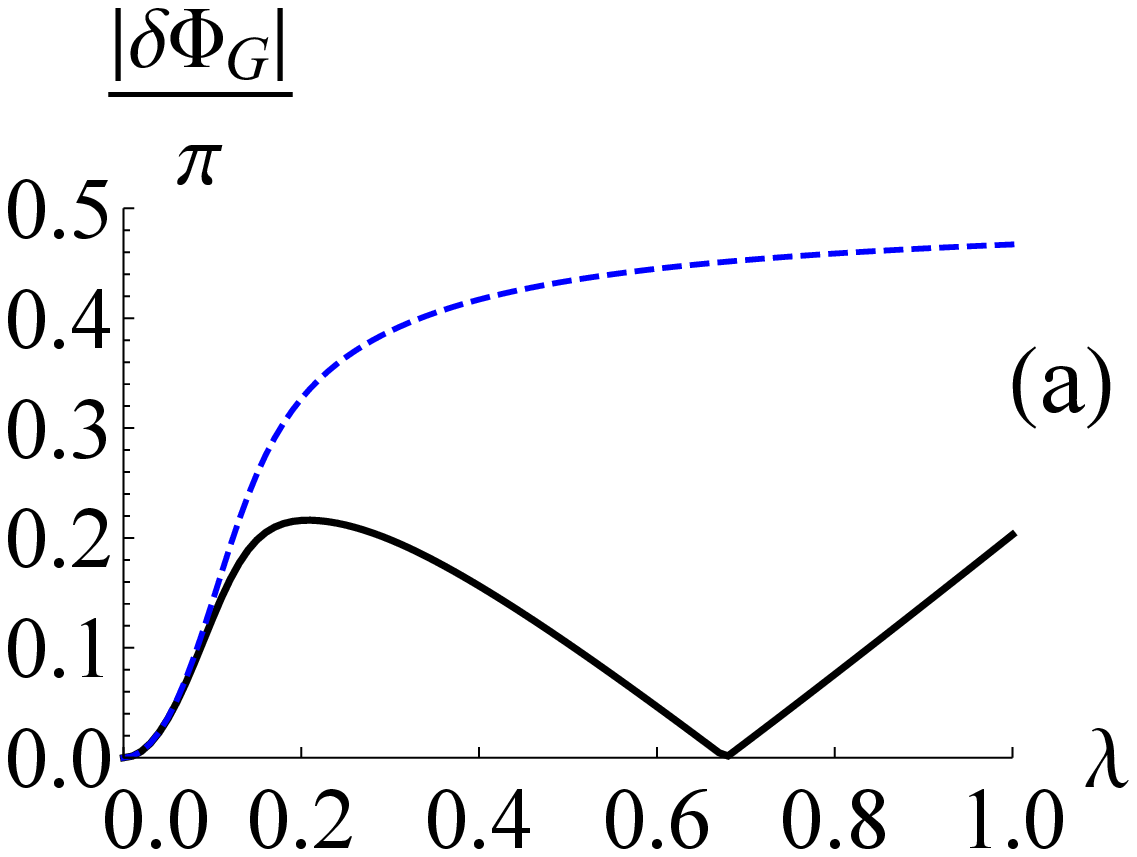}
\end{minipage}\hfill%
\begin{minipage}[t]{0.48\linewidth}
\includegraphics[width=\linewidth]{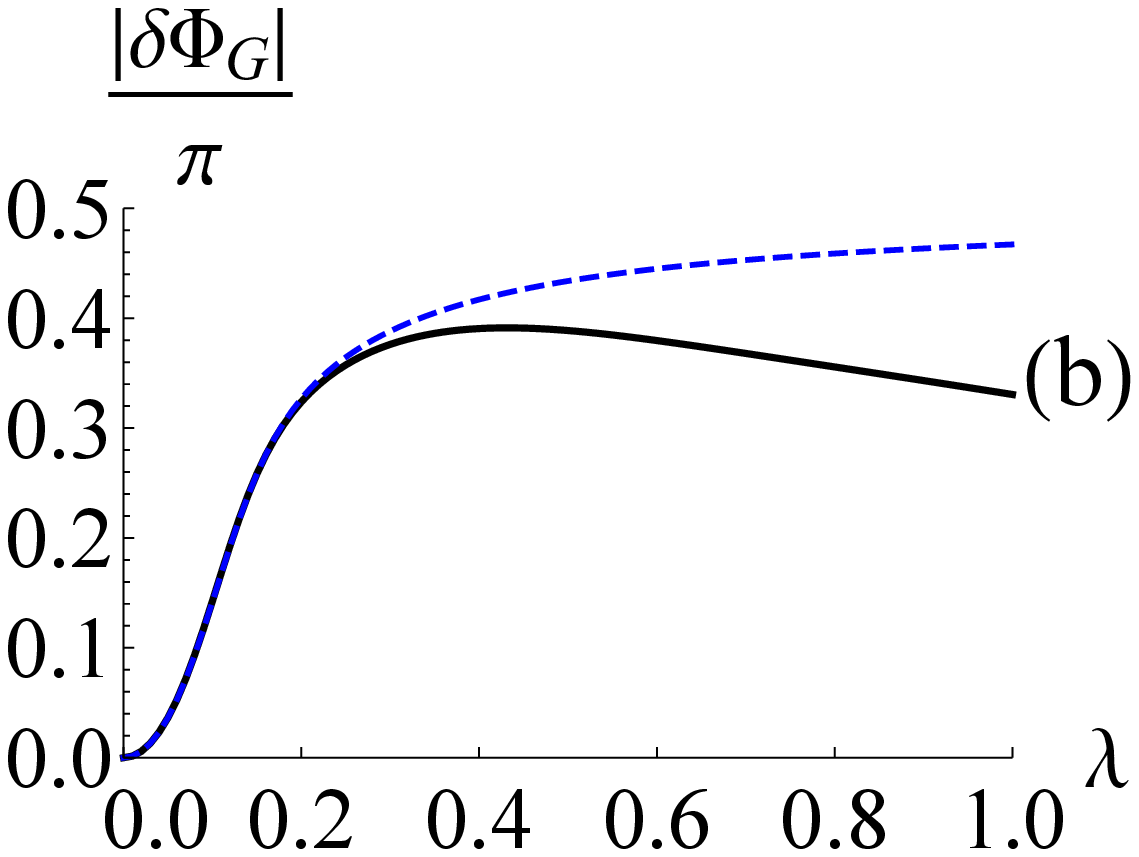}
\end{minipage}%
\caption{(Color online) Behavior of the correction to the geometric phase $\delta \Phi_G = \Phi_G - \Phi_U$ [here $\Phi_U = \pi (\cos \theta_0 - 1)$] as the spin-spin environment coupling strength $\lambda$ is varied, both with (solid, black) and without (dashed, blue) initial correlations when the initial state is prepared via a projective measurement. We have considered the environment to be a spin bath with $N = 50$, and for simplicity, we have assumed that the interaction strength between the central spin and each environment spin is the same (that is, $\lambda_j = \lambda$ for all $j$). As always, we are working in dimensionless units with $\hbar = 1$ and here have set $\omega_i=1$ for all $i$. In (a), we have used zero temperature $(\beta \rightarrow \infty)$, while in (b), $\beta = 0.4$. Also, $\omega_0 = 5$ and $\theta_0 = \pi/3$.}
\label{figspinenvprojprep}
\end{figure}

If the initial state is $\rho(0) = P_\psi e^{-\beta H} P_\psi/Z$, then the off-diagonal elements of the density matrix are given by 
\begin{align}
    \langle \sigma_{\pm}(t)\rangle&=\langle \sigma_{\pm}\rangle e^{\pm i[\omega_{0}t + \chi(t)]}e^{-\Gamma(t)}, 
\end{align}
where, similar to the form obtained for the harmonic oscillator environment, 
\begin{align*}
    \tan[\chi(t)]&=\frac{\sinh(\beta \omega_{0}/2)-\cos\theta_{0}\cosh(\beta\omega_{0}/2)}{\cosh(\beta \omega_{0}/2)-\cos\theta_{0}\sinh(\beta \omega_{0}/2)}\tan[\Phi(t)], 
\end{align*}
with $\theta_0$ the Bloch angle characterizing the initial state. We now have 
\begin{equation}
\Phi(t)=\sum_{j}\arg\big[A_{j}(t)+iB_{j}(t)\big],
\label{Phispin}
\end{equation}
where $A_{j}(t)=1- 2\frac{\lambda_{j}^2}{\alpha_{j^2}}\sin^2(\alpha_j t)$ and $B_{j}(t)=\frac{\lambda_{j}^2}{\alpha_{j}^2}\tanh(\beta \alpha_{j})\sin(2\alpha_j t)$ with $\alpha_{j}=\sqrt{\lambda_{j}^2+\omega_{j}^2}$. Also, $\Gamma(t) = \Gamma_{\text{uc}}(t) + \Gamma_{\text{corr}}(t)$, where $\Gamma_{\text{corr}}(t) =\Gamma_{\text{corr}}^{(1)}(t)+\Gamma_{\text{corr}}^{(2)}(t)$, and 
\begin{widetext}
\begin{align}
    \Gamma_{\text{corr}}^{(1)}(t)&=-\frac{1}{2}\sum_{j}\ln\bigg[1+(\lambda_j/\omega_j)^4\bigg(\frac{\tanh(\beta\alpha_j)\sin(2\alpha_jt)}{1+(\lambda_j/\omega_j)^2\cos(2\alpha_{j}t)} \bigg)^2\bigg],\label{gammaprojcorr1}\\
    \Gamma_{\text{corr}}^{(2)}(t)&= -\frac{1}{2}\ln\bigg[1-\frac{(1-\cos^2\theta_{0})\sin^2[\Phi(t)]}{[\cosh(\beta \omega_{0}/2)-\cos\theta_{0}\sinh(\beta\omega_{0}/2)]^2} \bigg]\label{gammaprojcorr2}.
\end{align}
\end{widetext} 
Interestingly, in this case, even if the temperature is zero, the initial correlations change the decay rate of the off-diagonal elements since $\Gamma_{\text{corr}}^{(1)}(t) \neq 0$ at zero temperature while $\Gamma_{\text{corr}}^{(2)}(t) = 0$. On the other hand, at zero temperature, $\chi(t)$ is once again equal to $\Phi(t)$.

With the system density matrix found, we compute the correction to the geometric phase $\delta \Phi_G = \Phi_G - \Phi_U$. The behavior of the correction $\delta \Phi_G$ as a function of the two-level system-environment coupling strength is shown in Figs.~\ref{figspinenvprojprep}(a) and (b). The effect of the initial correlations is again very significant; in particular, the initial correlations can make the geometric phase more robust. For particular values of the system-environment interaction strength $\lambda$, the correction to the geometric phase becomes zero.

\subsection{System state preparation by unitary operation}  

\begin{figure}[h!]
\begin{minipage}[t]{0.48\linewidth}
\includegraphics[width=.9\linewidth]{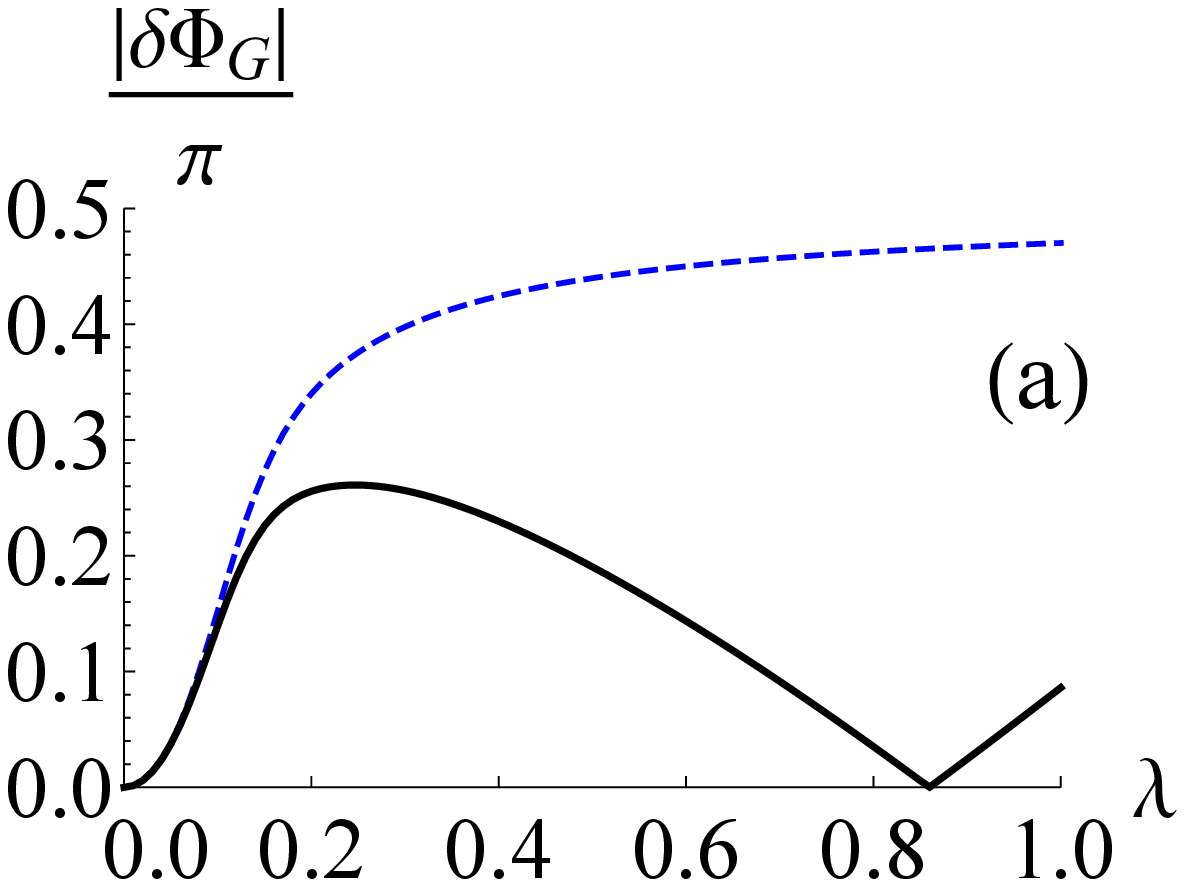}
\end{minipage}\hfill%
\begin{minipage}[t]{0.48\linewidth}
\includegraphics[width=.9\linewidth]{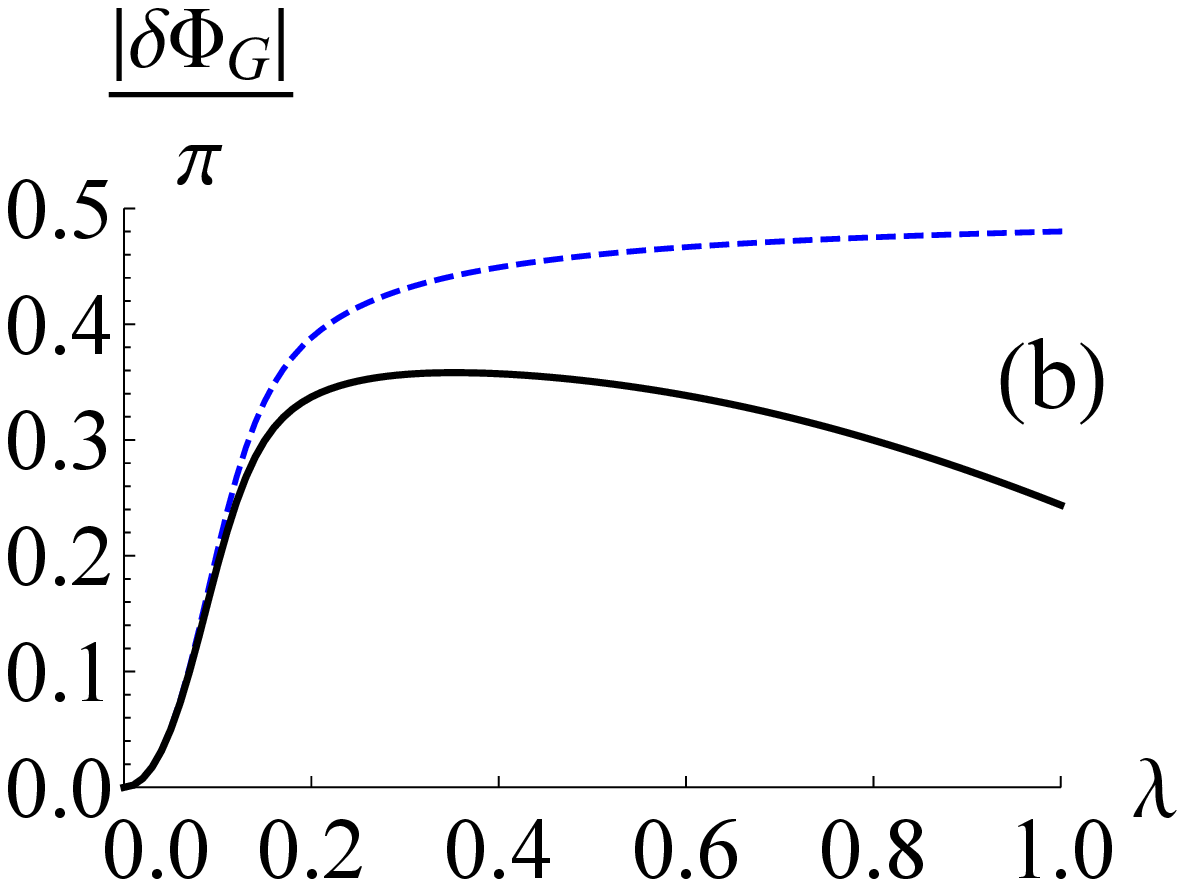}
\end{minipage}%
\caption{(Color online) Behavior of the correction to geometric phase $\delta \Phi_G = \Phi_G - \Phi_0$ (here $\Phi_0$ is the geometric phase when the system-environment coupling strength is zero) as the spin-spin environment coupling strength $\lambda$ is varied, both with (solid, black) and without (dashed, blue) initial correlations when the initial state is prepared by a unitary operation. As before, we have considered the environment to be a spin bath with $N = 50$, and $\lambda_j = \lambda$ for all $j$. Once again, we are working in dimensionless units with $\hbar = 1$ and here have set $\omega_i=1$ for all $i$. In (a), we have used $\beta = 1$, while in (b), $\beta = 0.4$. Also, $\omega_0 = 5$ and $\Omega = e^{i \pi \sigma_y/3}$.}
\label{spinenvunit}
\end{figure}

We now prepare the initial system state via a unitary operation. We find that for the initial system-environment state $\rho(0) = \frac{1}{Z}\Omega e^{-\beta H}\Omega^\dagger$, the off-diagonal elements of the density matrix are, as for the harmonic oscillator environment,
\begin{equation}
\langle \sigma_\pm (t) \rangle = \langle \sigma_\pm (0) \rangle e^{\pm i [\omega_0 t + \chi(t)]} e^{-\Gamma(t)},
\end{equation}    
where $\Gamma(t) = \Gamma_{\text{uc}}(t) + \Gamma_{\text{corr}}^{(1)}(t)+\Gamma_{\text{corr}}^{(2)}(t)$ with $\Gamma_{\text{corr}}^{(1)}(t)$ the same as before [see Eq.~\eqref{gammaprojcorr1}], while $\Gamma_{\text{corr}}^{(2)}(t)$ is given by 
\begin{widetext} 
\begin{align*}
 \Gamma_{\text{corr}}^{(2)}(t)&=-\ln \left\lbrace\text{abs}\left[\frac{e^{-\beta \omega_0/2}  \opav{0}{\Omega^\dagger \sigma_+ \Omega}{0} e^{- i\Phi(t)} + e^{\beta \omega_0/2}  \opav{1}{\Omega^\dagger \sigma_+ \Omega}{1} e^{ i\Phi(t)}}{e^{-\beta \omega_0/2}  \opav{0}{\Omega^\dagger \sigma_+ \Omega}{0}  + e^{\beta \omega_0/2}  \opav{1}{\Omega^\dagger \sigma_+ \Omega}{1}}\right]\right\rbrace. \notag\\
\end{align*}
\end{widetext}
Also, $\chi(t)$ is of the same form as in Eq.~\eqref{phicorrhounit}, but with $\Phi(t)$ now given by Eq.~\eqref{Phispin}. Details can be found in Appendix \ref{appendixspinenv}. Once again, for zero temperature, we find that the dynamics are the same as the case where the initial state is prepared by a projective measurement. However, as illustrated in Figs.~\ref{spinenvunit}(a) and (b), even for non-zero temperatures, the contribution to the geometric phase due to the initial correlations can be very significant. Once again, if we increase the temperature, the effect of the initial correlations decreases as expected.

\section{Conclusion}

In summary, we have presented exact expressions for the geometric phase of a two-level system undergoing pure dephasing to investigate the effect of the initial system-environment correlations on the geometric phase. As concrete examples, we have applied these expressions to two different environments: a collection of harmonic oscillators, and a collection of spins. Our results illustrate that the effect of the initial correlations on the geometric phase can be very significant, with a non-trivial dependence on the system-environment parameters. For instance, increasing the system-environment coupling strength may not always increase the correction to the geometric phase; in fact, for certain values of the coupling strength, the correction becomes zero, implying that the initial correlations can increase the robustness of the geometric phase. Our work on the geometric phase should be important not only for studies of the geometric phase itself as well as its practical implementations, but also for investigating the role of system-environment correlations in open quantum systems. 

\section*{acknowledgements}
The authors acknowledge support from the LUMS FIF Grant FIF-413. A.~Z.~C. is also grateful for support from HEC under grant No 5917/Punjab/NRPU/R\&D/HEC/2016. Support from the National Center for Nanoscience and Nanotechnology is also acknowledged.

\appendix

\section{Solution for harmonic oscillator environment}
\label{appendixhoenv}

For completeness, we sketch how to solve for the system dynamics for the total system-environment Hamiltonian $H = H_S + H_B + H_{SB}$, where \cite{MorozovPRA2012,ChaudhryPRA2013a} 
\begin{align*}
&H_S = \frac{\omega_0}{2} \sigma_z, \; H_B = \sum_k \omega_k b_k^\dagger b_k, \\
&H_{SB} = \sigma_{z}\sum_k (g_k^* b_k + g_k b_k^\dagger).
\end{align*}
First, we transform to the interaction picture to obtain
\begin{align}
H_I(t) &= e^{i(H_S + H_B)t} H_{\text{SB}} e^{-i(H_S + H_B)t}, \notag \\
&= \sigma_z \sum_k (g_k^* b_k e^{-i\omega_k t} + g_k b_k^\dagger e^{i\omega_k t} ).
\end{align}
We next find the time evolution operator $U_I(t)$ corresponding to $H_I(t)$ using the Magnus expansion as
\begin{equation}
U_I(t) = \exp \lbrace \sigma_z \sum_k [b_k^\dagger \alpha_k(t) - b_k \alpha_k^*(t)]/2\rbrace,
\end{equation}
and the total unitary time-evolution operator is $U(t) = e^{-i\omega_0 \sigma_z t/2} U_I(t)$. We now define $[\rho_S(t)]_{10} = \text{Tr}_{S,B} [U(t) \rho(0) U^\dagger (t) \ket{0}\bra{1}]$. Defining $P_{01}(t) = U^\dagger (t) \ket{0}\bra{1} U(t)$, this can be written as $[\rho_S(t)]_{10} = \text{Tr}_{S,B} [\rho(0) P_{01}(t)]$. Simplifying $P_{01}(t)$ using the unitary time-evolution operator $U(t)$, we find that 
\begin{equation}
P_{01}(t) = e^{i\omega_0 t} e^{-R_{01}(t)} P_{01},
\end{equation}
where
\begin{equation}
R_{01}(t) = \sum_k [b_k^\dagger \alpha_k(t) - b_k \alpha_k^*(t)],
\end{equation}
with 
$$\alpha_k(t) = \frac{2g_k (1 - e^{i\omega_k t})}{\omega_k}.$$
Consequently, 
\begin{eqnarray}
\label{generalrdmhoenv}
[\rho_S(t)]_{10}=  e^{i\omega_0 t} \text{Tr}_{S,B} [e^{-R_{01}(t)} P_{01} \rho(0)].
\end{eqnarray}
This is a general result because it applies to an arbitrary initial density $\rho(0)$. Now, if $\rho(0) = \rho_S(0) \otimes \rho_B$, where $\rho_B = \frac{e^{-\beta H_B}}{Z_B}$ with $Z_B = \text{Tr}_B [e^{-\beta H_B}]$, then
\begin{eqnarray}
[\rho_S(t)]_{10}  =  [\rho_S(0)]_{10} e^{i\omega_0t} \text{Tr}_B [e^{-R_{01}(t)} \rho_B].
\end{eqnarray}
The trace over the environment computes to 
\begin{align}
 &\text{Tr}_B[e^{-R_{01}(t)} \rho_B] =\notag \\
 &\exp \left[ -\sum_k  4 |g_k|^2 \frac{[1 - \cos (\omega_k t)]}{\omega_k^2} \coth \left(\frac{\beta \omega_k}{2} \right) \right],
\end{align}
thereby yielding 
\begin{equation}
[\rho_S(t)]_{10} = [\rho_S(0)]_{10} e^{i \omega_0 t}  e^{-\Gamma_{\text{uc}}(t)},
\end{equation}
with
\begin{equation}
\Gamma_{\text{uc}}(t) = \sum_k 4|g_k|^2 \frac{[1 - \cos (\omega_k t)]}{\omega_k^2} \coth \left( \frac{\beta \omega_k}{2} \right).
\end{equation}

We now consider what happens if the initial state is of the form $\rho(0) = \frac{1}{Z} \Omega e^{-\beta H} \Omega^\dagger$, with $Z$ the normalization factor. Currently, the $\Omega$ operator can be a projection operator or a unitary operator. To first simplify $Z$, we use the completeness relation $\sum_l \ket{l}\bra{l} = \mathds{1}$, where $\sigma_z \ket{l} = (-1)^l\ket{l}$. Then, 
\begin{align}
Z &= \sum_l e^{-\beta \omega_0 (-1)^l/2} \opav{l}{\Omega^\dagger \Omega}{l} \text{Tr}_B [e^{-\beta H_B^{(l)}}],
\end{align}
with 
\begin{equation}
H_B^{(l)} = H_B + (-1)^l \sum_k (g_k^* b_k + g_k b_k^\dagger).
\end{equation}
To simplify further, we introduce the displaced harmonic oscillator modes 
\begin{align}
B_{k,l} = b_k + \frac{(-1)^lg_k}{\omega_k}, \\
B_{k,l}^\dagger = b_k^\dagger + \frac{(-1)^lg_k^*}{\omega_k},
\end{align}
allowing us to write 
\begin{equation}
Z = \sum_l e^{-\beta \omega_0 (-l)^l/2} \opav{l}{\Omega^\dagger \Omega}{l} e^{\beta \sum_k \frac{|g_k|^2}{\omega_k}} Z_B,
\end{equation}
where $Z_B = \text{Tr}_B [e^{-\beta \sum_k \omega_k B_{k,l}^\dagger B_{k,l}} ]$. With $Z$ found, we then substitute our initial state in Eq.~\eqref{generalrdmhoenv} and introduce $\sum_l \ket{l}\bra{l}$ to simplify the resulting $\text{Tr}_B [ e^{-R_{01} (t)} e^{-\beta H_B^{(l)}} ]$. Using the displaced harmonic oscillator modes as before, we find that 
\begin{equation}
R_{01}(t) = \sum_k [\alpha_k (t) B_{k,l}^\dagger - \alpha_k^* (t) B_{k,l} ] + i(-1)^l\Phi(t),
\end{equation}
where
\begin{eqnarray}
\Phi(t) & = & \sum_k \frac{4|g_k|^2}{\omega_k^2} \sin (\omega_k t).
\end{eqnarray}
We then find that
\begin{equation}
\text{Tr}_B [e^{-R_{01}(t)} e^{-\beta H_B^{(l)}}] = e^{-i(-1)^l\Phi(t)}  Z_B e^{\beta \sum_k \frac{|g_k|^2}{\omega_k}} e^{-\Gamma_{\text{uc}}(t)}.
\end{equation} 
Putting this all together, and rearranging, we obtain 
\begin{align}
[\rho_S(t)]_{10} &= [\rho_S(0)]_{10} e^{i\omega_0 t}   e^{-\Gamma_{\text{uc}}(t)} X(t),
\end{align}
with 
$$X(t) = \dfrac{\sum_l  \opav{l}{\Omega^\dagger P_{01} \Omega}{l} e^{-i(-1)^l\Phi(t)} e^{-\beta \omega_0 (-1)^l/2} }{\sum_l \opav{l}{\Omega^\dagger P_{01} \Omega}{l} e^{-\beta \omega_0 (-1)^l/2} }.$$
Now assuming that $\Omega$ is a projection operator, that is, $\Omega = \ket{\psi}\bra{\psi}$, we can further simplify and write $X(t)$ in polar form to obtain Eq.~\eqref{hoenvprojectiveprep}. On the other hand, if $\Omega$ is taken to be a unitary operator, we obtain Eq.~\eqref{hoenvunitaryprep}.

\section{Dynamics with a spin environment}
\label{appendixspinenv}

We now consider the total system-environment Hamiltonian $H=H_S+H_B+H_{SB}$, where
$$ H_S = \frac{\omega_0}{2} \sigma_z, \; H_B =\sum_{i}   \omega_{i}\sigma^{i}_{x},\>\> H_{SB} = \sigma_z  \otimes \sum_i \lambda_i \sigma_z^i .$$
Once again, since $[H_S, H_{SB}] = 0$, this is a pure dephasing model. Our aim is to then calculate $\langle \sigma_{\pm}(t)\rangle$. We note that $e^{it(H_B+H_{SB})} \ket{l} = e^{it(H_B + (-1)^l V)}\ket{l}$, where
\begin{align}
    V=\sum_i \lambda_i \sigma_z^i. 
\end{align}
Using the completeness relation $\sum_s \ket{l}\bra{l} = \mathds{1}$, we can simplify $\sigma_{\pm}(t) = e^{iHt}\sigma_\pm e^{-iHt}$ to find 
\begin{align}
\label{sigmaplustime}
    \sigma_\pm(t)&=e^{ \pm i\omega_0 t}e^{it(H_B \pm V)}e^{-it(H_B \mp V)}\sigma_\pm.
\end{align}
We now consider initial states of the form
\begin{align}
    \rho(0)&=\rho_{S}(0)\otimes\rho_{B},\>\>\>\rho_{B}=e^{-\beta H_{B}}/Z_{B}.
\end{align}
For simplicity, we only show the calculation for $\langle \sigma_+(t) \rangle$. Using Eq.~\eqref{sigmaplustime}, we obtain 
\begin{align}
\label{sigg}
    &\langle \sigma_{+}(t)\rangle=\tr[\sigma_{+}(t)\rho(0)]\notag\\
    &=\frac{e^{i\omega_{0}t}}{Z_{B}} \langle \sigma_+(0) \rangle \tr_{B}[R(t)e^{-\beta H_{B}}],
\end{align}
where $R(t) = e^{it(H_{B}+V)}e^{-it(H_{B}-V)}$. Our remaining task is to compute $\tr_B [R(t) e^{-\beta H_B}]$. To this end, we first write $R(t)$ as $e^{it\sum_{j}\alpha_j (\vec{n}_{1}^{j}\cdot \vec{\sigma}_{j})}e^{-it\sum_{j}\alpha_j(\vec{n}_{2}^{j}\cdot \vec{\sigma}_{j})}$, where $\vec{n}_{1}^{j}=\frac{1}{\alpha_j}(\omega_{j}, 0, \lambda_{j})$, $\vec{n}_{2}^{j}=\frac{1}{\alpha_j}(\omega_{j}, 0, -\lambda_{j})$ and $\alpha_j = \sqrt{\omega_j^2 + \lambda_j^2}$. The exponentials can then be combined and the resulting expression is further simplified to obtain 
\begin{align}
    \tr_B [R(t) e^{-\beta H_B}] = 2^N \prod_j \cos c_j\cos(i\beta \omega_{j}),
\end{align}
where $\cos c_j=1-2\bigg(\frac{\lambda_{j}}{\alpha_j}\bigg)^{2}\sin^{2}(\alpha_j t)$, and $Z_{B}=2^N\Pi_{j}\cos(i\beta \omega_{j})$. Putting it all together, we finally have that 
\begin{align}
    \langle \sigma_{+}(t)\rangle= \langle \sigma_{+}(0)\rangle e^{i\omega_{0}t}\prod_{j}\bigg\{1-2\bigg(\frac{\lambda_{j}}{\alpha_j}\bigg)^{2}\sin^{2}(\alpha_j t) \bigg\}.
\end{align}

We now consider initially correlated states of the form
\begin{align*}
    \rho(0)&=\frac{1}{Z}\Omega_{}e^{-\beta H}\Omega^{\dagger}
\end{align*}
As before, we find that $\langle \sigma_{+}(t)\rangle=\tr_{S,B}[e^{i\omega_{0}t}R(t)\sigma_{+}\rho(0)]$. To simplify $\rho(0)$, we use the fact that $e^{-\beta H} \ket{s} =e^{(-1)^{s+1}\beta \omega_{0}/2}e^{-\beta(H_{B}+(-1)^s V)}\ket{s}$. We then have 
\begin{align}
    \langle \sigma_+(t)\rangle&=\frac{e^{i\omega_{0}t}}{Z}\bigg[\opav{0}{\Omega^{\dagger}\sigma_{+}\Omega}{0} e^{-\beta \omega_{0}/2} \tr_{B}[R(t)e^{-\beta(H_{B}+V)}]\notag \\&+\opav{1}{\Omega^{\dagger}\sigma_{+}\Omega}{1} e^{\beta \omega_{0}/2}\tr_{B}[R(t)e^{-\beta(H_{B}-V)}]\bigg].
\end{align}
We now sketch the calculation for $\tr_B [R(t) e^{-\beta(H_{B}+V)}]$ as the calculation for $\tr_{B}[R(t)e^{-\beta(H_{B}-V)}]$ is very similar. The trick is to write $\tr_B [R(t) e^{-\beta(H_{B}+V)}]$ as $\tr_B[e^{-it(H_{B}-V)}e^{i\gamma(H_{B}+V)}]$ we have defined $\gamma = t + i\beta$. The exponentials can then be manipulated as before to obtain 
\begin{align*}
\tr_{B}[R(t)e^{-\beta(H_{B}+V)}] = C_{0}\Pi_{j} \left(A_j - iB_j \right), 
\end{align*}
where 
$A_j = 1-2(\frac{\lambda_j}{\alpha_j})^{2}\sin^{2}(\alpha_j t)$, $B_j = 2(\frac{\lambda_{j}}{\alpha_j})^{2}\tanh(\beta \alpha_j)\sin(\alpha_j t) \cos(\alpha_j t)$, and $C_{0}=\Pi_{j}2\cosh(\beta \alpha_j)$. We can then further simplify to 
\begin{align*}
    \langle \sigma_{+}(t)\rangle = &\langle \sigma_+\rangle e^{i\omega_{0}t} e^{-\Gamma(t)} \times \\
    &\frac{\sum_l \opav{l}{\Omega^\dagger \sigma_+ \Omega}{l}e^{-\beta \omega_0 (-1)^l/2} e^{-i(-1)^l\Phi(t)}}{\sum_l \opav{l}{\Omega^\dagger \sigma_+ \Omega}{l}e^{-\beta \omega_0 (-1)^l/2}},
\end{align*}
where
\begin{align}
 \Gamma(t)&=\sum_{j}\Gamma_{j}(t), \>\>\>\Phi(t)=\sum_{j}\Phi_{j}(t),   
\end{align}
and $F_j(t)=A_{j}(t)+iB_{j}(t)=e^{-\Gamma_{j}(t)}e^{i\Phi_{j}(t)}$. It is then a simple matter of specifying that $\Omega$ is a projection operator or a unitary operator to work out the dynamics.

\end{document}